\documentclass{article}
\usepackage{graphicx}  
\usepackage{amsmath}   
\usepackage[compress]{cite}
\usepackage{amssymb}   
\usepackage{bm} 
\usepackage{dcolumn}
\usepackage{color}
\usepackage{mathrsfs}
\usepackage{amsfonts}
\RequirePackage[colorlinks,citecolor=blue,urlcolor=magenta,linkcolor=blue]{hyperref}
\def\eq#1{{Eq.~(\ref{#1})}}
\def\eqs#1{{Eqs.~(\ref{#1})}}
\def\sect#1{{Sec.~\ref{#1}}}
\def\EH{Einstein-Hilbert }
\def\LL{Lanczos-Lovelock }
\def\gr{general relativity}
\title{Lanczos-Lovelock gravity from a thermodynamic perspective}
\author{Sumanta Chakraborty
\footnote{sumanta@iucaa.ernet.in}
\footnote{sumantac.physics@gmail.com}\\
{\small {IUCAA, Post Bag 4, Ganeshkhind,}}\\
{\small {Pune University Campus, Pune 411 007, India}}}
\begin{document}
  
\maketitle
\begin{abstract}
The deep connection between gravitational dynamics and horizon thermodynamics leads to several intriguing features both in general relativity and in Lanczos-Lovelock theories of gravity. Recently in arXiv:1312.3253 several additional results strengthening the above connection have been established within the framework of general relativity. In this work we provide a generalization of the above setup to Lanczos-Lovelock gravity as well. To our expectation it turns out that most of the results obtained in the context of \gr\ generalize to Lanczos-Lovelock gravity in a straightforward \emph{but} non-trivial manner. First, we provide an alternative and more general derivation of the connection between Noether charge for a specific time evolution vector field and gravitational heat density of the boundary surface. This will lead to holographic equipartition for static spacetimes in Lanczos-Lovelock gravity as well. Taking a cue from this, we have introduced \emph{naturally} defined four-momentum current 
associated with gravity and matter energy momentum tensor for both Lanczos-Lovelock Lagrangian and its quadratic part. Then, we consider the concepts of Noether charge for null boundaries in Lanczos-Lovelock gravity by providing a direct generalization of previous results derived in the context of general relativity.

Another very interesting feature for gravity is that gravitational field equations for arbitrary static and spherically symmetric spacetimes with horizon can be written as a thermodynamic identity in the near horizon limit. This result holds in both general relativity and in Lanczos-Lovelock gravity as well. In a previous work [arXiv:1505.05297] we have shown that, for an arbitrary spacetime, the gravitational field equations near any null surface generically leads to a thermodynamic identity. In this work, we have \emph{also} generalized this result to Lanczos-Lovelock gravity by showing that gravitational field equations for Lanczos-Lovelock gravity near an arbitrary null surface can be written as a thermodynamic identity. Our general expressions under appropriate limits reproduce previously derived results for both the static and spherically symmetric spacetimes in Lanczos-Lovelock gravity. Also by taking appropriate limit to general relativity we can reproduce the results presented in arXiv:1312.3253 and 
arXiv:1505.05297.
\end{abstract}
\newpage
\tableofcontents
\newpage 
\section{Introduction}\label{Paper05:SecIntro}

In recent years, several interesting features have been derived for gravitational theories in which the field equations do not contain more than second order derivatives of the dynamical variable. This general class of gravity theories are known as \LL theories of gravity which includes \gr\ as a special case \cite{Lanczos1932,Lovelock1971,Padmanabhan2013b}. All these features are shown to stem from the deep connection between gravitational dynamics and horizon thermodynamics. Even though they first emerged in the context of \gr\ showing that Einstein's field equations near a horizon become a thermodynamic identity \cite{Padmanabhan2002,Padmanabhan2005a,Padmanabhan2010a,Cai2005,Akbar2006,Cai2007,Cai2008a,Akbar2007,Akbar2009,Gong2007,Wu2007,Padmanabhan2007b,Chakraborty2010,Chakraborty2009}, this result transcends general relativity and extends naturally to horizons in spherically symmetric and static spacetimes within \LL theories of gravity \cite{Padmanabhan2006a,Padmanabhan2009}. Horizons in general (and 
black holes in particular) possess thermodynamic attributes like entropy \cite{Bekenstein1973,Bekenstein1974} 
and temperature \cite{Hawking1975,Davies1976,Unruh1976,Hawking1977}. Also any null surface can act as a local Rindler horizon for some observer \cite{Padmanabhan2010a}. The above framework allows one to introduce observer-dependent thermodynamic variables near any arbitrary event in the spacetime. From all these evidences it seems natural to think of gravitational dynamics as a long wavelength thermodynamic limit of some microscopic degrees of freedom \cite{Jacobson1995,Padmanabhan2010b,Wald2001,Padmanabhan2008,Padmanabhan2002b,Padmanabhan2002c}. This emergent gravity paradigm has received significant amount of support from later investigations, especially from the following results: 
\begin{itemize}
\item 
The action functional for gravity is expressible as a sum of a bulk term and a surface term with a ``holographic" relation between them. This result holds both in Einstein gravity and in all \LL theories of gravity \cite{Padmanabhan2006c,Kolekar2010,Kolekar2012b}.
\item 
Gravitational field equations when projected on an arbitrary null surface, reduces to the Navier-Stokes equation of fluid dynamics in any spacetime \cite{Padmanabhan2011a,Kolekar2012a,Damour1982}.
\item
Gravitational field equations in all \LL\ models can be obtained from thermodynamic extremum principles \cite{Padmanabhan2007,Padmanabhan2008} involving the heat density of null surfaces in the spacetime.
\item 
In \cite{Krishna2013} a pair of conjugate variables $f^{ab}=\sqrt{-g}g^{ab}$ and $N^{c}_{ab}=-\Gamma ^{c}_{ab}+(1/2)(\Gamma ^{d}_{bd}\delta ^{c}_{a}+\Gamma ^{d}_{cd}\delta ^{a}_{b})$ have been introduced in terms of which the gravitational action can be interpreted as a momentum space action. Also variation of these variables has very natural thermodynamic interpretation, $\delta f^{ab}$ is related to variation of entropy and $\delta N^{c}_{ab}$ is related to variation of temperature when evaluated on an arbitrary null surface. This idea has been generalized to \LL gravity in \cite{Sumanta2014b} by introducing a new set of variables with identical thermodynamic interpretation.
\item
This paradigm also offers a possible solution to the cosmological constant problem. In this paradigm (a) the field equations are invariant under addition of a constant to the matter Lagrangian, (b) the cosmological constant appears as an integration constant and finally (c) its value can be determined by postulating a dimensionless number (known as CosMIn) to have a value $4\pi$ \cite{PadmanabhanH2013,PadmanabhanH2014}. This dimensionless number counts the number of modes that cross the Hubble volume at the end of inflation and re-enters the Hubble volume at the beginning of late-time acceleration phase \cite{PadmanabhanH2014}.   
\end{itemize}
Recently in \cite{Padmanabhan2013a} the connection between gravity and thermodynamics has been explored further in the context of general relativity. It has been demonstrated in \cite{Padmanabhan2013a} that, in the context of general relativity, the following results hold: 
\begin{itemize}
\item 
The total Noether charge contained in a 3-volume $\mathcal{R}$, evaluated for a specific time evolution vector field, can be interpreted as the surface heat density of the boundary $\partial \mathcal{R}$ of the volume.
\item 
The time evolution of the spacetime itself can be described by the difference between surface and bulk degrees of freedom. Here the surface degrees of freedom, $N_{\rm sur}$ is equal to the area of the boundary and $N_{\rm bulk}$ is the Komar energy modulo average Unruh-Davies temperature of the boundary surface. For static spacetime this will lead to holographic equipartition: $N_{\rm sur} = N_{\rm bulk}$. 
\item 
For suitably defined gravitational momentum related to a specific time evolution vector, total energy (gravity+matter) in a bulk volume $\mathcal{R}$ turns out to be the heat density of the surface $\partial \mathcal{R}$.
\item
For a bulk region bounded by null surfaces, the total Noether charge within that region is related to `heating' of the boundary surface.
\end{itemize}
In the past, virtually every result related to thermodynamic structure of gravity in the context of general relativity has been generalized to \LL models of gravity. In this case as well it is worth investigating whether the above description can also be generalized to \LL gravity. This is very important since the expression for horizon entropy in general relativity is just a quarter of the horizon area, while in \LL models, the corresponding expression is much more complex. Given this fact it is not clear a priori whether our results --- interpretation of Noether charge for both timelike and null surfaces along with gravitational momentum  ---  will generalize to \LL gravity. We will show in this work that, all these results possess a natural generalization to \LL gravity. 

There is another very important and curious connection between gravitational dynamics and horizon thermodynamics. This originates from the fact that field equations for gravity near a horizon in both static \cite{Padmanabhan2009} and spherically symmetric spacetime \cite{Padmanabhan2006a} can be written as a thermodynamics identity. In this work, we will try to generalize this result for an arbitrary spacetime with a null surface, which is neither static nor spherically symmetric. As we have mentioned earlier---by introducing local Rindler horizon it is possible to attribute thermodynamical entities like temperature and entropy to the null surface. Also components of energy momentum tensor has physical interpretation, e.g., the $T^{r}_{r}$ component (where $r$ is like the radial coordinate) can be taken as a measure of radial pressure in spherically symmetric spacetime. Then we can consider an infinitesimal displacement of the null surface along the outgoing null geodesic $k^{a}$ with affine parameter $\
lambda$. From the above virtual displacement we can ask the physical meaning of the following object $T\delta _{\lambda}S-P\delta _{\lambda}V$. In both static and spherically symmetric spacetimes the above object is $\delta _{\lambda}E$, for some suitably defined energy $E$. Thus through this exercise it would be possible to identify the \emph{energy} (the relation of this object with standard notions of energy will be discussed elsewhere) for an arbitrary null surface through its appearance in the thermodynamic identity. Starting from this thermodynamic identity in an arbitrary spacetime with a null surface, taking suitable limits we can arrive at both the static and spherically symmetric results respectively. Thereby verifying previous results along this direction in the literature explicitly from our general formulation. 

The paper is organized as follows: First in \sect{Paper05:Sec:LLIntro} we have provided a brief introduction to \LL gravity to make the reader familiar with notations and conventions. Next in \sect{Paper05:Sec:02} we will derive the equivalence of gravitational Field equation with a thermodynamic identity for an arbitrary null surface, which then is applied in \sect{Paper05:Sec:03} to static spacetime and spherically symmetric spacetime to retrieve the respective result in those cases as well. Furthermore, in \sect{Paper05:Sec:LLGen} we have presented all the results generalized to \LL gravity, which involves Noether charge, gravitational momentum and null surfaces. Finally, we have concluded with a discussion on our results. All the relevant derivations are summarized in App. \ref{Paper05:AppA}, App. \ref{Paper05:AppB} and App. \ref{Paper05:AppC}. 

Throughout this paper, we work in a $D$ dimensional spacetime where we use metric signature $\left(-,+,+,+,\ldots\right)$ with all the fundamental constants $G$, $\hbar$ and $c$ being set to unity. All the Latin indices a,b,$\ldots$ run from $0$ to $(D-1)$, Greek indices $\mu$, $\nu$, $\ldots$ run from $1$ to $(D-1)$ and capitalized Latin indices A,B,$\ldots$ stand for transverse coordinates.

\section{A Brief Introduction to \LL Gravity}\label{Paper05:Sec:LLIntro}

In this paper, we will work exclusively within the framework of \LL gravity. Thus before going to the main body of this work it is advantageous to introduce some definitions and notations that we will follow throughout \cite{Padmanabhan2013b}. For that purpose we will provide a brief introduction to the \LL gravity itself. 

We will start by consider the most general setup of a D-dimensional spacetime in which the action functional for gravity is described by an arbitrary function of metric $g^{ab}$ and curvature $R^{a}_{~bcd}$. Thus the action functional takes the following form:
\begin{equation}\label{Paper05:SecLL:01}
A=\int _{\mathcal{V}}d^{D}x \sqrt{-g}L\left(g^{ab},R^{a}_{~bcd}\right)
\end{equation}
It should be emphasized that the gravitational Lagrangian introduced above depends both on the curvature and the metric but not on the derivatives of the curvature i.e., the Lagrangian contains terms only up to second order derivatives of the metric. The most important quantity, which will be extremely helpful for our later purpose, derived from the Lagrangian, is the following tensor:
\begin{equation}\label{Paper05:SecLL:02}
P^{abcd}=\left(\frac{\partial L}{\partial R_{abcd}} \right)_{g_{ij}}
\end{equation}
This tensor has all the algebraic symmetry properties of the curvature tensor, namely: (a) antisymmetry in first two and last two indices, (b) symmetry under pair exchange and finally (c) cyclic identity. Using this tensor an analogue of the Ricci tensor in \gr\ can be constructed by the following definition 
\begin{equation}\label{Paper05:SecLL:03}
\mathcal{R}^{ab}\equiv P^{aijk}R^{b}_{~ijk}.
\end{equation}
This tensor is indeed symmetric under interchange of the indices $(a,b)$ alike the Ricci tensor; but the result is nontrivial to prove (for this result and other properties of these tensors, see \cite{Padmanabhan2011a}). Having defined the action functional the next obvious task is to consider a variation of the action functional to get the field equations. The variation of the action presented in \eq{Paper05:SecLL:01} leads to the result:
\begin{equation}\label{Paper05:SecLL:04}
\delta A=\int _{\mathcal{V}}d^{D}x \sqrt{-g}E_{ab}\delta g^{ab}
+\int _{\mathcal{V}}d^{D}x \sqrt{-g}\nabla _{j}\delta v^{j}
\end{equation}
where $E_{ab}$ represents the field equation term resulting from the variation of the bulk part of the action and $\delta v^{a}$ is the boundary term which we generally set to zero at the boundaries (if the boundary term contains normal derivatives then we need to add a counter term to the action). Both the field equation term and the boundary term are given by the following expressions:
\begin{subequations}
\begin{align}\label{Paper05:SecLL:05}
E_{ab}&\equiv \mathcal{R}_{ab}-\frac{1}{2}g_{ab}L-2\nabla ^{m}\nabla ^{n}P_{amnb}
\\
\delta v^{j}&=2P^{ibjd}\nabla _{b}\delta g_{di}-2\delta g_{di}\nabla _{c}P^{ijcd}
\end{align}
\end{subequations}
The quantity $P^{abcd}$ involves second order derivatives of the metric which in turn implies that the term $\nabla ^{m}\nabla ^{n}P_{amnb}$ in $E^{ab}$ contains fourth order derivatives of the metric. Therefore, in order to get field equation containing only second order derivatives of the metric  we must impose an extra condition on $P^{abcd}$, such that
\begin{equation}\label{Paper05:SecLL:06}
\nabla _{a}P^{abcd}=0.
\end{equation}
Hence the problem of obtaining field equation having up to second order derivatives of the metric from an action functional reduces to that of finding scalar functions of curvature and metric such that \eq{Paper05:SecLL:06} is satisfied. It turns out that such a scalar indeed exists and is unique; it is given by the \LL Lagrangian \cite{Lovelock1971,Padmanabhan2010b,Padmanabhan2013b,Padmanabhan2006a,Padmanabhan2006c} in D dimensions, as 
\begin{equation}\label{Paper05:SecLL07}
L=\sum _{m}c_{m}L_{m}=\sum _{m}\frac{c_{m}}{m}
\frac{\partial L_{m}}{\partial R_{abcd}}R_{abcd}=\sum _{m}\frac{c_m}{m}P^{abcd}_{(m)}R_{abcd}
\end{equation}
The Lagrangian $L_{m}$ is a homogeneous function of $R_{abcd}$ of order $m$ which can be written as $L_{m}=Q^{abcd}_{(m)}R_{abcd}$. This can be used to identify $P^{abcd}_{(m)}=mQ^{abcd}_{(m)}$. From now on we shall work with this $m$th order Lagrangian since any result obtained for $L_{m}$ can be generalized to most general Lagrangian in \eq{Paper05:SecLL07} in a straightforward manner. Henceforth we shall drop the $m$ index. For this $m$th order \LL Lagrangian we have the following explicit expression for $P_{cd}^{ab}$ in terms of the curvature tensor:
\begin{equation}\label{Paper05:SecLL08}
P^{ab}_{cd}=\frac{\partial L_{m}}{\partial R^{cd}_{ab}}
=\frac{m}{16\pi}\frac{1}{2^{m}}\delta ^{aba_{2}b_{2}\ldots a_{m}b_{m}}_{cdc_{2}d_{2}\ldots c_{m}d_{m}}
R^{c_{2}d_{2}}_{a_{2}b_{2}}\ldots R^{c_{m}d_{m}}_{a_{m}b_{m}}\equiv mQ^{ab}_{cd}
\end{equation}
where we have $\delta ^{aba_{2}b_{2}\ldots a_{m}b_{m}}_{cdc_{2}d_{2}\ldots c_{m}d_{m}}$ to be completely antisymmetric determinant tensor. This relation will be used extensively later. 

Diffeomorphism invariance is a property that any generally covariant theory shares, including the \LL theories of gravity. This has serious implications; the invariance of action functional under an infinitesimal coordinate transformation, $x^{a}\rightarrow x^{a}+\xi ^{a}(x)$ lead to conservation of a current, usually called the \emph{Noether current}. From variation of action functional we can get the Noether current having the following expression \cite{Padmanabhan2013b,Padmanabhan2010b}:
\begin{equation}\label{Paper05:SecLL09}
J^{a}\equiv \left( 2E^{ab}\xi _{b}+L\xi ^{a}+\delta _{\xi}v^{a}\right)
\end{equation}
(However in a recent work \cite{Padmanabhan2013a} it has been shown that the conservation of Noether current follows from identities in differential geometry alone; which has been generalized for \LL gravity in \cite{Sumanta2014a}.) In the above expression the last term, $\delta _{\xi}v^{a}$ represents the boundary term of the \LL Lagrangian. From the conservation property of the Noether current i.e., $\nabla _{a}J^{a}=0$, we can define an antisymmetric tensor known as \emph{Noether Potential} as, $J^{a}=\nabla _{b}J^{ab}$. General expressions for these quantities can be found in \cite{Padmanabhan2010b}. In the context of \LL theories, we have $\nabla _{a}P^{abcd}=0$, and the Noether current and Noether potential simplifies to (for most general case see \cite{Chakraborty2014c})
\begin{subequations}
\begin{align}
J^{ab}&=2P^{abcd}\nabla _{c}\xi _{d}
\label{Paper05:SecLL10a}
\\
J^{a}&=2P^{abcd}\nabla _{b}\nabla _{c}\xi _{d}
=2\mathcal{R}^{ab}\xi _{b}+2P_{k}^{~ija}\pounds _{\xi}\Gamma ^{k}_{ij}
\label{Paper05:SecLL10b}
\end{align}
\end{subequations}
where $\Gamma ^{a}_{bc}$ is the metric compatible connection. 

A direct thermodynamic interpretation can be presented for the Noether current. For that one requires to associate Wald entropy with horizons in all \LL models. The corresponding entropy density (which, integrated over the horizon gives the entropy) is given by \cite{Padmanabhan2010a,Padmanabhan2012b,Wald1993,
Strominger1996,Ashtekar1998,Bombelli1986}
\begin{eqnarray}\label{Paper05:SecLL11}
s=-2\pi \sqrt{q} P^{abcd}\mu _{ab}\mu _{cd}
\end{eqnarray}
where $\mu _{ab}$ is the bi-normal to the $(D-2)$-dimensional surface and $q$ is the respective determinant. The best way to see that this is indeed the entropy density is to consider the \EH limit. In which case we have $P^{ab}_{cd}=(1/32\pi)(\delta ^{a}_{c}\delta ^{b}_{d}-\delta ^{a}_{d}\delta ^{c}_{b})$ and $\mu _{ab}=(u_{a}r_{b}-r_{a}u_{b})$ with $u^{2}=-1$ and $r^{2}=+1$ leading to $s=\sqrt{q}/4$.

\section{Equivalence of Gravitational Field Equation Near an Arbitrary Null Surface to Thermodynamic Identity}\label{Paper05:Sec:02}

An arbitrary spacetime with a null surface can always be parametrized using Gaussian Null Coordinates (henceforth refereed to as GNC). This coordinate system can be constructed in analogy with Gaussian normal coordinates. In the non-null case the construction proceeds by invoking geodesics normal to the desired surface. While for a null surface characterized by null normal $\ell _{a}$ the normal geodesics are on the surface. Then construction of coordinates \emph{off} the surface can be achieved by introducing an auxiliary null vector $k^{a}$ satisfying $\ell _{a}k^{a}=-1$ and moving away from the null surface along the null geodesics of $k^{a}$. The construction of such a coordinate system for an arbitrary spacetime with a null surface has been detailed in \cite{Moncrief1983,Morales2008,Parattu2015a} and the line element in this $D$-dimensional spacetime turns out to have the following form:
\begin{equation}\label{Paper05:Sec:02:Eq.01}
ds^{2}=-2r\alpha du^{2}+2dudr-2r\beta _{A}dudx^{A}+q_{AB}dx^{A}dx^{B}
\end{equation}
where $x^{A}$'s are the $(D-2)$ transverse coordinates. Note that the above line element contains $(D-2)(D-1)/2$ independent parameters in $q_{AB}$, $(D-2)$ independent parameters in $\beta _{A}$ and finally one independent parameter $\alpha$. All of them are dependent on the coordinates $\left(u,r,x^{A}\right)$. The surface $r=0$ is the null surface under our consideration. The null normal $\ell _{a}$ and the auxiliary vector $k_{a}$ has the following expressions \cite{Parattu2015a}:
\begin{subequations}
\begin{align}
\ell _{a}&=\left(0,1,0,0\right),\qquad \ell ^{a}=\left(1,2r\alpha +r^{2}\beta ^{2},r\beta ^{A}\right)
\label{Paper05:Sec:02:Eq.02a}
\\
k_{a}&=\left(-1,0,0,0\right),\qquad k^{a}=\left(0,-1,0,0\right)
\label{Paper05:Sec:02:Eq.02b}
\end{align}
\end{subequations}
It turns out that, for the $r=0$ surface the non-affinity parameter corresponding to the null normal $\ell ^{a}$ is obtained from $\ell ^{b}\nabla _{b}\ell ^{a}=\kappa \ell ^{a}$. This yields the non-affinity parameter to be: $\kappa =\alpha$. While the vector $k_{a}=-\nabla _{a}u$, is tangent to the ingoing null geodesic and is affinely parametrized, with affine parameter $r$. Hence on the null surface we denote $\lambda _{H}$ to be the value of the affine parameter. Also in the remaining discussion we will work with $\lambda$ being identified as: $\lambda -\lambda_{H}=r$.

The $m$th order \LL Lagrangian in $D$ dimension is given by (throughout this paper we will follow the notation of \cite{Padmanabhan2013b,Padmanabhan2007b,Padmanabhan2009}):
\begin{align}\label{Paper05:Sec:02:Eq.03}
\mathcal{L}_{m}&=\frac{1}{16\pi}\frac{1}{2^{m}}\delta ^{a_{1}b_{1}\ldots a_{m}b_{m}}_{c_{1}d_{1}\ldots c_{m}d_{m}}R^{c_{1}d_{1}}_{a_{1}b_{1}}\cdots R^{c_{m}d_{m}}_{a_{m}b_{m}}
\end{align}
A general \LL Lagrangian consists of a linear combination of various $m$ terms with different coefficients. However a result if true for a given order $m$, will hold for the most general linear combination, follows immediately. Thus we will restrict ourselves by considering a $m$th order \LL Lagrangian (in the literature it has often been argued that pure Lovelock is more fundamental than the total \LL action itself, see e.g., \cite{Dadhich2015,Dadhich2012,Dadhich2010}), from which the field equation turns out to be (we will drop the subscript $m$ from now on):
\begin{align}\label{Paper05:Sec:02:Eq.04}
E^{i}_{j}&=-\frac{1}{2}\frac{1}{16\pi}\frac{1}{2^{m}}\delta ^{ia_{1}b_{1}\ldots a_{m}b_{m}}_{jc_{1}d_{1}\ldots c_{m}d_{m}}R^{c_{1}d_{1}}_{a_{1}b_{1}}\ldots R^{c_{m}d_{m}}_{a_{m}b_{m}}
\nonumber
\\
&=\frac{1}{16 \pi}\frac{m}{2^{m}}\delta ^{a_{1}b_{1}\ldots a_{m}b_{m}}_{jd_{1}\ldots c_{m}d_{m}}R^{id_{1}}_{a_{1}b_{1}}\ldots R^{c_{m}d_{m}}_{a_{m}b_{m}}-\frac{1}{2}\delta ^{i}_{j}\mathcal{L}
=\frac{1}{2}T^{i}_{j}
\end{align}
The equivalence of the two expressions in the first and second line follows from \cite{Padmanabhan2013b,Padmanabhan2009}. In GNC coordinates we will now consider the near null surface behavior of gravitational field equation in the $m$th order \LL gravity. As in Einstein gravity \cite{Chakraborty2015a} in this case as well we will start with a subclass of the GNC parametrization in order to bring out the physics involved. For that as in the case of \EH action in \LL gravity as well we will impose two additional requirements, namely $\beta _{A}\vert _{r=0}=0$ and hypersurface orthogonality for time-like unit vector $u_{a}$ constructed from $\xi _{a}$. This immediately leads to $\partial _{A}\alpha \vert _{r=0}=0$ (see \eq{Paper05:AppB:Eq.02a} in Appendix \ref{Paper05:AppB}). Hence these two conditions implies that $\alpha$ should be independent of transverse coordinates. This can be thought of as an \emph{extension of the zeroth law of black hole thermodynamics for an arbitrary null surface in \LL theories of 
gravity}. On using this condition we arrive at the following expression for $T^{ab}\ell _{a}k_{b}$ (which equals $T^{r}_{r}$ in the null limit) as (see \eq{Paper05:AppA:05new} in App. \ref{Paper05:AppA}):
\begin{align}\label{Paper05:Sec:02:Eq.05}
T^{r}_{r}&=2E^{r}_{r}=\frac{m}{8}\frac{1}{2^{m-1}}\left(\frac{\alpha}{2\pi}\right)\left(\delta ^{PA_{1}B_{1}\ldots A_{m-1}B_{m-1}}_{QC_{1}D_{1}\ldots C_{m-1}D_{m-1}}R^{C_{1}D_{1}}_{A_{1}B_{1}}\ldots R^{C_{m-1}D_{m-1}}_{A_{m-1}B_{m-1}}\right)\left(q^{QE}\partial _{r}q_{PE}\right)
\nonumber
\\
&-\frac{1}{16\pi}\frac{1}{2^{m}}\delta ^{A_{1}B_{1}\ldots A_{m}B_{m}}_{C_{1}D_{1}\ldots C_{m}D_{m}}R^{C_{1}D_{1}}_{A_{1}B_{1}}\ldots R^{C_{m}D_{m}}_{A_{m}B_{m}}
\nonumber
\\
&-\frac{m}{8\pi}\frac{1}{2^{m}}\Big\lbrace \delta ^{PA_{1}B_{1}\ldots A_{m-1}B_{m-1}}_{QC_{1}D_{1}\ldots C_{m-1}D_{m-1}}\Big[-\frac{1}{2}q^{QE}\partial _{u}\partial _{r}q_{PE}
+\frac{1}{4}\left(q^{QE}\partial _{r}q_{PF}\right)\left(q^{FL}\partial _{u}q_{EL}\right)\Big]
\nonumber
\\
&\times R^{C_{1}D_{1}}_{A_{1}B_{1}}\ldots R^{C_{m-1}D_{m-1}}_{A_{m-1}B_{m-1}}+2(m-1)\delta ^{PA_{1}B_{1}\ldots A_{m-1}B_{m-1}}_{QC_{1}D_{1}\ldots C_{m-1}D_{m-1}}R^{QC_{1}}_{uP}R^{uD_{1}}_{A_{1}B_{1}}\ldots R^{C_{m-1}D_{m-1}}_{A_{m-1}B_{m-1}}\Big\rbrace
\end{align}
Let us now consider the \EH limit of the above equation, which can be obtained by substituting $m=1$ in the above equation. This leads to:
\begin{align}\label{Paper05:Sec:02:Eq.06}
T^{r}_{r}=-\delta ^{A}_{B}R^{uB}_{uA}-\frac{1}{4}\delta ^{AB}_{CD}R^{CD}_{AB}
\end{align}
which exactly coincides the expression obtained in \cite{Chakraborty2015a}. Hence our general result reduces to the corresponding one in \EH action under appropriate limit.  

Now let us multiply \eq{Paper05:Sec:02:Eq.05} with the virtual displacement along $k^{a}$, which is $\delta \lambda$, and $\sqrt{q}$, where $q$ is the determinant of the transverse metric leading to:
\begin{align}\label{Paper05:Sec:02:Eq.07}
T^{r}_{r}\delta \lambda \sqrt{q}
&=\frac{m}{8}\frac{1}{2^{m-1}}\sqrt{q}\left(\frac{\alpha}{2\pi}\right)\left(\delta ^{PA_{1}B_{1}\ldots A_{m-1}B_{m-1}}_{QC_{1}D_{1}\ldots C_{m-1}D_{m-1}}R^{C_{1}D_{1}}_{A_{1}B_{1}}\ldots R^{C_{m-1}D_{m-1}}_{A_{m-1}B_{m-1}}\right)\left(q^{QE}\delta _{\lambda}q_{PE}\right)
\nonumber
\\
&-\delta \lambda \sqrt{q} \Big\lbrace \frac{1}{16\pi}\frac{1}{2^{m}}\delta ^{A_{1}B_{1}\ldots A_{m}B_{m}}_{C_{1}D_{1}\ldots C_{m}D_{m}}R^{C_{1}D_{1}}_{A_{1}B_{1}}\ldots R^{C_{m}D_{m}}_{A_{m}B_{m}}+\frac{m}{8\pi}\frac{1}{2^{m-1}}\delta ^{PA_{1}B_{1}\ldots A_{m-1}B_{m-1}}_{QC_{1}D_{1}\ldots C_{m-1}D_{m-1}}
\nonumber
\\
&\times \Big[-\frac{1}{2}q^{QE}\partial _{u}\partial _{r}q_{PE}
+\frac{1}{4}\left(q^{QE}\partial _{r}q_{PF}\right)\left(q^{FL}\partial _{u}q_{EL}\right)\Big]R^{C_{1}D_{1}}_{A_{1}B_{1}}\ldots R^{C_{m-1}D_{m-1}}_{A_{m-1}B_{m-1}}
\nonumber
\\
&+\frac{m(m-1)}{8\pi}\frac{1}{2^{m-1}}\delta ^{PA_{1}B_{1}\ldots A_{m-1}B_{m-1}}_{QC_{1}D_{1}\ldots C_{m-1}D_{m-1}}R^{QC_{1}}_{uP}R^{uD_{1}}_{A_{1}B_{1}}\ldots R^{C_{m-1}D_{m-1}}_{A_{m-1}B_{m-1}}
\Big\rbrace 
\end{align}
Now $\alpha /2\pi$ can be interpreted as the temperature associated with the null surface, and using $\delta _{\lambda}s$ from \eq{Paper05:AppA:09} we can integrate the above equation over $(D-2)$ dimensional null surface yielding (the most general expression has been provided in App. \ref{Paper05:AppA}):
\begin{align}\label{Paper05:Sec:02:Eq.08}
\int d\Sigma T^{r}_{r}\delta \lambda &=T\delta _{\lambda}\int d^{D-2}x~s-\int d\Sigma \delta \lambda\Big[\left(\frac{1}{16\pi}\frac{1}{2^{m}}\delta ^{A_{1}B_{1}\ldots A_{m}B_{m}}_{C_{1}D_{1}\ldots C_{m}D_{m}}R^{C_{1}D_{1}}_{A_{1}B_{1}}\ldots R^{C_{m}D_{m}}_{A_{m}B_{m}}\right)
\nonumber
\\
&+\Big\lbrace \frac{m}{8\pi}\frac{1}{2^{m-1}}\delta ^{PA_{1}B_{1}\ldots A_{m-1}B_{m-1}}_{QC_{1}D_{1}\ldots C_{m-1}D_{m-1}}\Big[-\frac{1}{2}q^{QE}\partial _{u}\partial _{r}q_{PE}
+\frac{1}{4}\left(q^{QE}\partial _{r}q_{PF}\right)\left(q^{FL}\partial _{u}q_{EL}\right)\Big]
\nonumber
\\
&\times R^{C_{1}D_{1}}_{A_{1}B_{1}}\ldots R^{C_{m-1}D_{m-1}}_{A_{m-1}B_{m-1}}+\frac{m(m-1)}{8\pi}\frac{1}{2^{m-1}}\delta ^{PA_{1}B_{1}\ldots A_{m-1}B_{m-1}}_{QC_{1}D_{1}\ldots C_{m-1}D_{m-1}}R^{QC_{1}}_{uP}R^{uD_{1}}_{A_{1}B_{1}}\ldots R^{C_{m-1}D_{m-1}}_{A_{m-1}B_{m-1}}
\Big\rbrace \Big]
\end{align}
where $d\Sigma =d^{D-2}x\sqrt{q}$. To bring out the physics presented in \eq{Paper05:Sec:02:Eq.08} we introduce the concept of transverse metric $g_{ab}^{\perp}$ and a work function \cite{Kothawala2010,Hayward1997}. Let us start by considering $u_{a}$ to be a normalized timelike vector and another normalized but spacelike vector $r_{a}$. They are related to the null vectors ($\ell_a,k_a$) by the following relations: $u_{a}=(1/2A)\ell _{a}+Ak_{a}$ and $r_{a}=(1/2A)\ell _{a}-Ak_{a}$, where $A$ is an arbitrary function. Given this setup the transverse metric is defined as $g_{ab}^{\perp}=u_{a}u_{b}-r_{a}r_{b}=\ell_{a}k_{b}+\ell_{b}k_{a}$. Using this transverse metric the work function is defined \cite{Kothawala2010,Hayward1997} to be $P=(1/2)T_{ab}g^{ab}_{\perp}=T_{ab}\ell ^{a}k^{b}$. In the adapted coordinate system on the null surface using the null vectors from \eqs{Paper05:Sec:02:Eq.02a} and (\ref{Paper05:Sec:02:Eq.02b}) we have $P=T^{r}_{r}$. As an aside we would like to mention that in the case of 
spherically symmetric spacetime, $P$ will be the transverse pressure. However, in this work we will not bother to illustrate the physical meaning of $P$ which can be obtained from \cite{Kothawala2010,Hayward1997}. 

Given this physical input we can rewrite \eq{Paper05:Sec:02:Eq.08} in the following form:
\begin{align}\label{Paper05:Sec:02:Eq.New01}
\bar{F}\delta \lambda =T\delta _{\lambda}S-\delta _{\lambda}E
\end{align}
This exactly coincides with the conventional first law of thermodynamics (not the first law for black hole mechanics but that in conventional thermodynamics, for some related works in black hole thermodynamics see \cite{Kastor2009,Dolan2011,Mann2012,Mann2014}), provided: (i) we identify the quantity $S$ to be the entropy of the null surface in \LL gravity and this \emph{exactly} matches with existing expression for entropy in \LL gravity \cite{Jacobson1993,Clunan2004,Wald1993}. (ii) We interpret $\bar{F}$ to be the average force over the null surface which is defined as the integral of the work function over the null surface as
\begin{align}\label{Paper05:Sec:02:Eq.New02}
\bar{F}=\int d^{D-2}x\sqrt{q}P
\end{align}
Finally, (iii) we should identify the second term on the right hand side of the \eq{Paper05:Sec:02:Eq.New01} as variation of an energy as the null surface is moved by an affine parameter distance $\delta \lambda$. The energy variation due to motion of the null surface has the following expression:
\begin{align}\label{Paper05:Sec:02:Eq.09}
\delta _{\lambda}E&=\delta \lambda \int d\Sigma \Big\lbrace \frac{1}{16\pi}\frac{1}{2^{m}}\delta ^{A_{1}B_{1}\ldots A_{m}B_{m}}_{C_{1}D_{1}\ldots C_{m}D_{m}}R^{C_{1}D_{1}}_{A_{1}B_{1}}\ldots R^{C_{m}D_{m}}_{A_{m}B_{m}}
\nonumber
\\
&+\frac{m}{8\pi}\frac{1}{2^{m-1}}\delta ^{PA_{1}B_{1}\ldots A_{m-1}B_{m-1}}_{QC_{1}D_{1}\ldots C_{m-1}D_{m-1}}\Big[-\frac{1}{2}q^{QE}\partial _{u}\partial _{r}q_{PE}
+\frac{1}{4}\left(q^{QE}\partial _{r}q_{PF}\right)\left(q^{FL}\partial _{u}q_{EL}\right)\Big]
\nonumber
\\
&\times R^{C_{1}D_{1}}_{A_{1}B_{1}}\ldots R^{C_{m-1}D_{m-1}}_{A_{m-1}B_{m-1}}+\frac{m(m-1)}{8\pi}\frac{1}{2^{m-1}}\delta ^{PA_{1}B_{1}\ldots A_{m-1}B_{m-1}}_{QC_{1}D_{1}\ldots C_{m-1}D_{m-1}}R^{QC_{1}}_{uP}R^{uD_{1}}_{A_{1}B_{1}}\ldots R^{C_{m-1}D_{m-1}}_{A_{m-1}B_{m-1}}
\Big\rbrace 
\end{align}
In the above expression for energy the $\lambda$ integral must be done \emph{after} the above integral has been performed, which in turn tells us that the detailed form of the expression inside bracket is required to get $E$ explicitly. 

The thermodynamic identity obtained in \eq{Paper05:Sec:02:Eq.New01} can be better explained if written in the following fashion:
\begin{align}
\delta _{\lambda}E=T\delta _{\lambda}S-\bar{F}\delta \lambda
\end{align}
This equation in this context can be interpreted as: Work done due to infinitesimal virtual displacement from $r=0$ to $r=\delta \lambda$ of the null surface, subtracted from the heat energy, i.e., temperature times entropy change, equals to the energy engulfed during this process. It is to be noted that in the general relativistic limit the last term in the energy expression would be absent and the second term in it leads to time rate of change of transverse area. Hence the energy expression for general relativity can be obtained by taking suitable limit of the above energy expression.
\section{Applications}\label{Paper05:Sec:03}

In the previous section we have derived the equivalence of gravitational field equations in \LL gravity to the thermodynamic identity $P\delta _{\lambda}V=T\delta _{\lambda}S-\delta _{\lambda}E$ near an arbitrary null surface. In this section we will illustrate two applications of our general result: First, the case of an arbitrary static spacetime (see \cite{Padmanabhan2009}). Second, the spherically symmetric spacetime (see \cite{Padmanabhan2006a}). The discussion will be brief since the details are sketched in the references cited above.  
\subsection{Stationary spacetime}

A spacetime will be called stationary, when we impose Killing conditions on the time evolution vector field. In GNC the most natural time evolution vector field corresponds to: $\boldsymbol{\xi}=\partial /\partial u$. Imposing Killing condition on this vector demands all the metric components, namely $\alpha$, $\beta _{A}$ and $q_{AB}$ to be independent of the $u$ coordinate (see \ref{Paper05:AppB}). Thus imposing this condition the energy expression as given in \eq{Paper05:Sec:02:Eq.09} reduces to the following form:
\begin{align}\label{Paper05:Sec:03:Eq.New09}
\delta _{\lambda}E&=\delta \lambda \int d\Sigma \Big\lbrace \frac{1}{16\pi}\frac{1}{2^{m}}\delta ^{A_{1}B_{1}\ldots A_{m}B_{m}}_{C_{1}D_{1}\ldots C_{m}D_{m}}R^{C_{1}D_{1}}_{A_{1}B_{1}}\ldots R^{C_{m}D_{m}}_{A_{m}B_{m}}
\Big\rbrace 
\nonumber
\\
&=\delta r\int d\Sigma ~\mathcal{L}^{(D-2)}
\end{align}
which immediately leads to the following differential equation for energy:
\begin{align}
\dfrac{\partial E}{\partial r}=\int d\Sigma ~\mathcal{L}^{(D-2)}
\end{align}
This exactly matches with the expression given in \cite{Padmanabhan2009}. For $m=1$ and $D=4$ this reduces to the expression obtained for Einstein's gravity. 

In order to achieve staticity we must impose hypersurface orthogonality on the time evolution vector $\xi ^{a}$ (or, equivalently on the four velocity constructed out of it). As mentioned in \ref{Paper05:AppB} this requires $\partial _{A}\alpha \vert _{r=0}$ to vanish. However from \eq{Paper05:Sec:03:Eq.New09} it is evident that this leads to no modification to our energy expression.

This is an interesting result. It shows that in arriving at this relation we have used two assumptions, viz, (a) $\beta _{A}=0$ on the null surface and (b) the spacetime is stationary. Hence the above result does not require spacetime to be static. Thus starting from the thermodynamic identity $T\delta _{\lambda}S=\delta _{\lambda}E+P\delta _{\lambda}V$ for arbitrary null surface we have shown that it holds for arbitrary static and stationary spacetimes as well. We will now take up the case for spherically symmetric but time dependent spacetime. 

\subsection{Spherically symmetric Spacetime}

We will finish by considering another application of our result: a spherically symmetric but \emph{not} necessarily static spacetime. GNC metric can be expressed in a spherically symmetric form by choosing the transverse coordinates $x^{A}$ to be the angular coordinates and enforcing $(D-2)$-sphere geometry on the $(u,r)=\textrm{constant}$ surface. Then we have the following restrictions on the GNC parameters, namely, $\partial _{A}\alpha =0$, $\beta _{A}=0$ and $q_{AB}=f(u,r)d\Omega _{(D-2)}^{2}$. When these conditions are imposed the line element takes the following form:
\begin{align}
ds^{2}=-2r\alpha (r,u)du^{2}+2dudr+f(u,r)d\Omega _{(D-2)}^{2}
\end{align}
We will define the radial coordinate \cite{Chakraborty2015a} as: $R(r,u)=\sqrt{f(r,u)}$. Then making a Taylor series expansion about $r=0$, we get $R(r,u)=R_{H}(u)+rg(r,u)$. Hence the null surface has a radius $R_{H}(u)$, which can change with $u$. This clearly shows that we have spherical symmetry but have retained time dependence. Hence by imposing spherical symmetry the energy satisfies a partial differential equation with the following form:
\begin{align}
\dfrac{\partial E}{\partial \lambda}&=\int d\Sigma \Big\lbrace \frac{1}{16\pi}\frac{1}{2^{m}}\delta ^{A_{1}B_{1}\ldots A_{m}B_{m}}_{C_{1}D_{1}\ldots C_{m}D_{m}}R^{C_{1}D_{1}}_{A_{1}B_{1}}\ldots R^{C_{m}D_{m}}_{A_{m}B_{m}}
\nonumber
\\
&+\frac{m}{8\pi}\frac{1}{2^{m-1}}\delta ^{PA_{1}B_{1}\ldots A_{m-1}B_{m-1}}_{QC_{1}D_{1}\ldots C_{m-1}D_{m-1}}\Big[-\frac{1}{2}q^{QE}\partial _{u}\partial _{r}q_{PE}
+\frac{1}{4}\left(q^{QE}\partial _{r}q_{PF}\right)\left(q^{FL}\partial _{u}q_{EL}\right)\Big]
\nonumber
\\
&\times R^{C_{1}D_{1}}_{A_{1}B_{1}}\ldots R^{C_{m-1}D_{m-1}}_{A_{m-1}B_{m-1}}+\frac{m(m-1)}{8\pi}\frac{1}{2^{m-1}}\delta ^{PA_{1}B_{1}\ldots A_{m-1}B_{m-1}}_{QC_{1}D_{1}\ldots C_{m-1}D_{m-1}}
R^{QC_{1}}_{uP}R^{uD_{1}}_{A_{1}B_{1}}\ldots R^{C_{m-1}D_{m-1}}_{A_{m-1}B_{m-1}}
\Big\rbrace 
\end{align}
A much simpler form can be derived in which the 2-sphere line element is just $(r+R_{H})^{2}d\Omega _{(D-2)}^{2}$, where the radial coordinate at the null surface $R_{H}$ is a constant. Then trading off $r$ in favor of $R$ the line element becomes
\begin{align}
ds^{2}=-2\left(R-R_{H}\right)\alpha (R,u)du^{2}+2dudR+R^{2}d\Omega _{(D-2)}^{2}
\end{align}
Then using the result that $\partial _{u}R_{H}=0$, the differential equation for energy can be integrated leading to,
\begin{align}\label{Newequationenergy}
E(R,u)&=\int d\lambda \int d\Sigma ~\mathcal{L}^{(D-2)}+X(u)
\nonumber
\\
&=\frac{1}{16\pi}A_{D-2}R ^{D-(2m+1)}\prod _{j=2}^{2m}(D-j)+X(u)
\end{align}
where $X(u)$ is an arbitrary function appearing as a ``constant'' of integration and $A_{D-2}$ originates from the differential volume element. Having introduced the radial coordinate $R$, we can replace $\lambda$ by $R$, since $R-R_{H}=r$, which coincides with the defining equation for $\lambda$. As we move along ingoing radial lines, which are also ingoing radial null geodesic $-\partial /\partial r$, we will gradually hit the center of the $(D-2)$-spheres (assuming that it exists). The affine parameter at the center would be $R=0$. Then \eq{Newequationenergy} at the center turns out to be:
\begin{align}
E(R=0,u)=X(u)
\end{align}
Since the center is a single point it is natural to associate zero energy with it, which determines the arbitrary function to be $X(u)=0$. Thus substituting this result in \eq{Newequationenergy} and evaluating on the null surface, we obtain the energy associated with the null surface in a spherically symmetric spacetime to be given by 
\begin{align}\label{Paper05:Sec:03:Eq.07}
E=\frac{1}{16\pi}A_{D-2}R_{H}^{D-(2m+1)}\prod _{j=2}^{2m}(D-j)
\end{align}
This again matches exactly with the result obtained in \cite{Padmanabhan2006a}. Another important thing to note is the following: the expression for energy, first obtained for static spherically symmetric configuration \emph{also} holds for time dependent situation with \emph{only} spherical symmetry assumed. 

\section{Various Geometrical Quantities in \LL Gravity and Their Thermodynamic Interpretation}\label{Paper05:Sec:LLGen}

As we have mentioned earlier, in this work we will be dealing exclusively with \LL gravity. For that purpose we have provided in \sect{Paper05:Sec:LLIntro} a short and brief introduction to the main aspects of the \LL gravity. As prescribed in the beginning we will generalize all the results presented in \cite{Padmanabhan2013a} to \LL gravity. Even though the concepts like Noether charge, Noether current for \LL gravity are well known in the literature and have been discussed at some length in \cite{Sumanta2014a} and  \cite{Sumanta2014b} in the connection with thermodynamic perspectives. Nevertheless we will first present an alternative derivation of the results related to Noether current. Then we will provide a generalization of a four vector from Einstein gravity to \LL gravity, which will carry the notion of gravitational momentum. Then we will concentrate on variation of this momentum and its meaning in thermodynamic language. The same steps will be followed for another momentum defined using only the 
quadratic part of the \LL action. Finally we discuss the null surfaces in the context of \LL gravity.

\subsection{The Spacetime Foliation}\label{Paper05:Sec:LL:Setup}

We will work in a spacetime which is being foliated by a series of spacelike hypersurfaces. These hypersurfaces are determined by the constant value of a scalar field $t(x)$. The unit normal to the $t(x)=\textrm{constant}$ hypersurface is given by $u_{a}=-N\nabla _{a}t$. If we consider $t$ as another coordinate then the above four-vector reduces to $-N\delta ^{0}_{a}$. For this spacetime foliation (which is known as slicing) we have $g^{00}=-1/N^{2}$, and the timelike normal being unit normalized i.e. $u^{a}u_{a}=-1$. Given this foliation, it is possible to introduce a time evolution vector field $\zeta ^{a}$ by the condition $\zeta ^{a}\nabla _{a}t=1$. In this coordinate system with $t$ as a coordinate it reduces to $\zeta ^{a}=\delta ^{a}_{0}$. This suggests the following decomposition: $\zeta ^{a}=-\left(\zeta ^{b}u_{b}\right)u^{a}+N^{a}$, where we have $N^{a}u_{a}=0$ and $N^{a}=h^{a}_{b}\zeta ^{b}$. In the above discussion the object $h^{a}_{b}=\delta ^{a}_{b}+u^{a}u_{b}$ represents the induced metric on 
the $t=\textrm{constant}$ surface. The above decomposition also introduces another vector 
\begin{equation}\label{Paper05:Setup:Eq.01}
\xi _{a}=Nu_{a}\to-N^{2}\delta ^{0}_{a}
\end{equation}
where the last result holds in the preferred foliation with $t$ as a coordinate. If we further impose the condition that $g_{0\alpha}=0$ this vector reduces to $\zeta ^{a}$. Also, in static spacetimes $\xi ^{a}$ turns out to be the time-like Killing vector field. $\xi ^{a}$ was shown to play a key role in ref. \cite{Padmanabhan2013a} to arrive at the thermodynamic interpretation. This vector also provides a rich structure as far as the Noether current and spacetime dynamics is concerned. 

\subsection{Noether Current and Related Aspects}\label{Paper05:Sec:LL:Noether}

Gravitational dynamics can be very efficiently described by using conserved current, known as Noether current. In literature this conserved current is shown to originate from the diffeomorphism invariance of the action. However this pose an immediate conceptual difficulty. In Electrodynamics for example, the gauge transformation is a symmetry of the system and thus is connected to a conserved current, which is trivial. Along identical lines the diffeomorphism invariance of gravitational action is also a gauge symmetry. Thus motivated by the example of electrodynamics it seems natural to expect the Noether current to originate from some differential geometry identity. This has been achieved recently in \cite{Padmanabhan2013a} in the context of \gr\ and subsequently was shown to hold in \LL gravity as well in \cite{Sumanta2014a}. Both these approaches show that the Noether current associated with a vector field $v^{a}$ can be obtained \emph{without} any use of diffeomorphism invariance of gravitational action. 
The proof involves connecting antisymmetric part of $\nabla _{i}v_{j}$ with $\mathcal{R}^{i}_{j}v^{j}$. From which the relation $\nabla _{a}J^{a}(v)=0$ follows just as an identity in differential geometry.

The striking feature of the above result is that a conserved current obtained from differential geometry turns out to be the one originating from diffeomorphism invariance of the action. In this work we will try to show the important role played by Noether current from a thermodynamic perspective, which can possibly shed light on this remarkable feature.

\paragraph*{Noether Charge and Surface Heat Content}

As emphasized earlier the vector $\xi ^{a}=Nu^{a}$ plays a central role in this work. Thus we start by computing Noether charge for the vector $\xi ^{a}$. Since this has been discussed extensively in \cite{Padmanabhan2013a,Sumanta2014a} we will be brief in this discussion. However we will point out some peculiarities which have not been noticed in earlier works. Firstly in all the previous works it has been assumed that the tensor $P_{a}^{~bcd}$ (see \eq{Paper05:SecLL:02}) has all the symmetry properties of the curvature tensor. However as we have shown explicitly in App. \ref{Paper05:AppC:Noether} the derivation of the Noether current itself does not require all the symmetry properties. We only need antisymmetry of $P^{abcd}$ in the pairs $(a,b)$ and $(c,d)$ along with pair exchange symmetry between them and vanishing divergence in the first two indices. Hence the same form of Noether current continues to hold for even general class of Lagrangians for which $P^{abcd}$ does not follow cyclic identity and is 
not divergence free in the last two indices. Secondly, in both the previous derivations the Noether current for $\xi ^{a}$ has been derived using the property that $J^{a}(v)=0$ where $v_{a}=\nabla _{a}\phi$. However in this work we have provided an altogether different derivation of the Noether current for $\xi ^{a}$. In the derivation we have used the bi-normals to $N=\textrm{constant}$ surface within the $t=\textrm{constant}$ surface. This leads to:
\begin{align}\label{Paper05:LL:Noether:Eq.01}
J^{ab}(\xi)=-2NaP^{abcd}\epsilon _{cd};\qquad J^{ab}(u)=-aP^{abcd}\epsilon _{cd}
\end{align}
where $a$ is the magnitude of the acceleration four vector obtained as: $a^{i}=u^{j}\nabla _{j}u^{i}$. Also in the above expressions we have: $\epsilon _{ab}=(u_{a}\hat{a}_{b}-u_{b}\hat{a}_{a})$ to be the bi-normal constructed out of $u_{a}$ and unit vector along acceleration, i.e., $\hat{a}_{i}$. The above two results suggest the following relation between Noether potential for the vector field $u_{a}$ and $\xi _{a}$:
\begin{align}\label{Paper05:LL:Noether:Eq.02}
J^{ab}(\xi)=2NJ^{ab}(u)
\end{align} 
Then the Noether current for $\xi ^{a}$ and hence the Noether charge can be obtained in a straightforward manner by differentiating the above equation, which leads to (see App. \ref{Paper05:AppC:Noether}):
\begin{align}\label{Paper05:LL:Noether:Eq.03}
u_{a}J^{a}(\xi)&=Nu_{a}J^{a}(u)+\nabla _{b}\left(N\chi ^{b}\right)
\nonumber
\\
&=\nabla _{b}\left(N\chi ^{b}\right)+ND_{\alpha}\chi ^{\alpha}=2D_{\alpha}\left(N\chi ^{\alpha}\right)
\end{align}
In arriving at the last line we have used the following results: $N\nabla _{i}\chi ^{i}=D_{\alpha}(N\chi ^{\alpha})$ and $D_{\alpha}\chi ^{\alpha}=\nabla _{i}\chi ^{i}-a_{i}\chi ^{i}$. The vector $\chi ^{i}$ is analogue of the acceleration four-vector in \gr\ and has the following expression:
\begin{align}\label{Paper05:LL:Noether:Eq.04}
\chi ^{b}=2P^{bacd}u_{a}u_{c}a_{d}
\end{align}
which the properties: $u_{a}\chi ^{a}=0$ and in the \gr\ limit $\chi ^{a}\rightarrow a^{a}$. The next natural object to consider is the total Noether charge, which can be obtained by integrating \eq{Paper05:LL:Noether:Eq.03} with integration measure $\sqrt{h}d^{D-1}x$. This leads to:
\begin{align}\label{Paper05:LL:Noether:Eq.05}
\int _{\mathcal{R}}\sqrt{h}d^{D-1}x~u_{a}J^{a}(\xi)=\int _{\partial \mathcal{R}}d^{D-2}x~\sqrt{q}\left(2Nr_{a}\chi ^{a} \right)
\end{align}
where $r_{a}$ represents the unit normal to the $N=\textrm{constant}$ surface within the $t=\textrm{constant}$ surface. This normal is equal to $\epsilon a_{a}/a$, where $\epsilon =\pm 1$ when the normal is directed towards the acceleration or vice versa. Then the above relation using \eq{Paper05:LL:Noether:Eq.04} can be written as:
\begin{align}\label{Paper05:LL:Noether:Eq.06}
\int _{\mathcal{R}}\sqrt{h}d^{D-1}x~u_{a}J^{a}(\xi)=\epsilon \int _{\partial \mathcal{R}}d^{D-2}x~\left(\frac{Na}{2\pi}\right)\left(8\pi P^{abcd}r_{a}u_{b}u_{c}r_{d}\right)=\epsilon \int _{\partial \mathcal{R}} T_{\rm loc}s
\end{align}
where $s$ is the entropy density obtained from \eq{Paper05:SecLL11} and $T_{\rm loc}$ corresponds to Tolman red-shifted local Unruh-Davies temperature as measured by an observer with four-velocity $u_{a}=-N\delta ^{0}_{a}$. Hence the total Noether charge inside the volume $\mathcal{R}$ corresponds to the heat density $Ts$ integrated over the boundary surface $\partial \mathcal{R}$.

\paragraph*{Holographic Equipartition} 

The above result showing the connection between Noether charge and heat content can be extended and interpreted in a completely different form. Using \eq{Paper05:SecLL10b} the Noether current can be expressed in terms of Lie variation of the connection and Lovelock Ricci $\mathcal{R}_{ab}$ as:
\begin{align}\label{Paper05:LL:Noether:Eq.07}
2u_{a}P_{i}^{~jka}\pounds _{\xi}\Gamma ^{i}_{jk}=D_{\alpha}\left(2N\chi ^{\alpha}\right)-N\bar{T}_{ab}u^{a}u^{b}
\end{align}
where we have used the relation: $\mathcal{R}_{ab}=(1/2)\bar{T}_{ab}=(1/2)\left[T_{ab}-(1/(D-2m))g_{ab}T \right]$. Then using the definition of Komar energy density as: $\rho _{Komar}=2N\bar{T}_{ab}u^{a}u^{b}$ and integrating over $(D-1)$ dimensional space $\mathcal{R}$ we arrive at \cite{Sumanta2014a}:
\begin{align}\label{Paper05:LL:Noether:Eq.08}
\int _{\mathcal{R}} \sqrt{h}d^{D-1}x~4u_{a}P_{i}^{~jka}\pounds _{\xi}\Gamma ^{i}_{jk}&=\epsilon \left(\frac{1}{2}T_{\rm avg}\right)\left(N_{\rm sur}-N_{\rm bulk} \right)
\end{align}
where $T_{\rm avg}$ corresponds to average of $Na/2\pi$, the local Unruh-Davies temperature of the observer with four-velocity $u_{a}$ over the boundary surface. $N_{\rm sur}$ encodes the surface degrees of freedom which equals temperature average of $4S$ and $N_{\rm bulk}$ determines the bulk degrees of freedom defined as Komar energy modulo $(1/2)T_{\rm avg}$. Note that the left hand side differs from the result obtained in \cite{Padmanabhan2013a,Sumanta2014a} by a factor of $16\pi$. This is due to the fact that the $P^{abcd}$ in those references differs from our convention by precisely a factor of $1/16\pi$. 

The above equation is actually equivalent to the field equation for \LL gravity but provides a `holographic' interpretation \footnote{In this work the word ``holography'' has been used in a primitive sense i.e. implying a relation between surface and bulk properties. This should not be confused with frequent use of this word in the context of string theory.}. The left hand side involving Lie variation of the connection $\Gamma ^{a}_{bc}$ provides time evolution of the spacetime which is solely dependent on the difference between suitably defined bulk and surface degrees of freedom. Note that for static spacetime $\xi ^{a}=\delta ^{a}_{0}$ and hence is the Killing vector. Then Lie variation of $\Gamma ^{a}_{bc}$ along the Killing vector $\xi ^{a}$ vanishes. This immediately leads to:
\begin{align}\label{Paper05:LL:Noether:Eq.09}
N_{\rm sur}=N_{\rm bulk}
\end{align}
Thus for static spacetime surface degrees of freedom coincides with the bulk degrees of freedom. This  provides the holographic equipartition between surface and bulk degrees of freedom. When holographic equipartition does not hold the difference between surface and bulk degrees of freedom is responsible for evolution of the spacetime in \LL gravity. 

\subsection{Bulk Gravitational Dynamics And Its Relation to Surface Thermodynamics in \LL Gravity}
\label{Paper05:Sec:LL:Bulk}

In \cite{Padmanabhan2013a} it has been illustrated that, total energy of matter and gravity equals the surface heat content in the context of \gr. To prove this a suitably defined gravitational four-momentum $P^{a}$ has been used such that when integrated over a $t=\textrm{constant}$ surface with proper integration measure it leads to a notion of gravitational energy. The notion of energy is quiet ambiguous in the sense of observer dependence. For example, even in special relativity the energy of a particle with four momentum $\mathbf{p}$ as measured by an observer with four velocity $\mathbf{u}$ is: $E=-\mathbf{u}.\mathbf{p}$. This immediately suggests that we should use identical trick to identify the energy by contracting a suitably defined four momentum $P^{a}$ with the four velocity $u_{a}$ introduced in \sect{Paper05:Sec:LL:Setup}. We will first briefly describe the situation for \gr\ and shall generalize subsequently to \LL gravity.

\paragraph*{Bulk Energy Versus Surface Heat Energy}

The \EH action can be written explicitly in terms of two canonically conjugate variables, namely, $f^{ab}=\sqrt{-g}g^{ab}$ and $N^{c}_{ab}=Q^{cp}_{aq}\Gamma ^{q}_{bp}+Q^{cp}_{bq}\Gamma ^{q}_{ap}$ \cite{Krishna2013}. It turns out that the \EH action can be interpreted as a momentum space action with $f^{ab}$ as the coordinate and $N^{c}_{ab}$ as its conjugate momentum \cite{Krishna2013}. Using these two variables a natural definition of gravitational momentum can be provided as \footnote{Note that the negative sign in front of $P^{a}$ is just a convention, such that $-u_{a}P^{a}$ relates to the energy.}:
\begin{align}\label{Paper05:LL:Momentum:Eq.01}
-P^{a}(q)=g^{ij}\pounds _{q}N^{a}_{ij}+q^{a}L
\end{align}
where $q^{a}$ is an arbitrary vector field. This follows from the result that Hamiltonian can be written as $H=q\dot{p}+L$ with the identification of $p$ as $N^{c}_{ab}$ and $q$ as $g^{ab}$ (or $f^{ab}$ if we consider $-\sqrt{-g}P^{a}$). However in order to generalize the above result to \LL gravity, we should rewrite the above expression in a slightly modified form, such that: $g^{ij}\pounds _{q}N^{a}_{ij}=2g^{ij}\pounds _{q}(Q^{ap}_{im}\Gamma ^{m}_{jp})=2Q_{m}^{~jpa}\pounds _{q}\Gamma ^{m}_{jp}$. Thus the above action can also be interpreted with $\Gamma ^{m}_{jp}$ and $Q_{m}^{~jpa}$ as conjugate variables. Surprisingly, this is also a valid pair of conjugate variables as illustrated in \cite{Sumanta2014b} and can be generalized readily to \LL gravity. However in this case we need to interpret $\Gamma ^{m}_{jp}$ as the momenta and $Q_{m}^{~jpa}$ as the coordinate, which is true since $\Gamma ^{m}_{jp}$ has 40 independent degrees of freedom while $Q_{m}^{~jpa}$ as constructed out of the metric only has 10 
independent degrees of freedom (a detailed discussion has been presented in \cite{Sumanta2014b}). The above setup can be generalized in a natural fashion to \LL gravity following \cite{Sumanta2014b} and leads to:
\begin{align}\label{Paper05:LL:Momentum:Eq.02}
-\sqrt{-g}P^{a}(q)=2\sqrt{-g}P_{p}^{~qra}\pounds _{q}\Gamma ^{p}_{qr}
+\sqrt{-g}Lq^{a}
\end{align} 
The physical structure of this momentum can be understood in greater detail by using the Noether current. Writing the corresponding expression for Noether current explicitly in the case of \LL gravity and then simplifying we obtain (see App. \ref{Paper05:AppC:EHMomentum}):
\begin{align}\label{Paper05:LL:Momentum:Eq.03}
J^{a}(q)=\nabla _{b}J^{ab}(q)=2E^{a}_{b}q^{b}-P^{a}(q)
\end{align}
This leads to another definition for the momentum which will turn out to be quiet useful and can be given as,
\begin{align}\label{Paper05:LL:Momentum:Eq.04}
-P^{a}(q)=\nabla _{b}J^{ab}(q)-2E^{a}_{b}q^{b}
\end{align}
Then divergence of the momentum has the following expression:
\begin{align}\label{Paper05:LL:Momentum:Eq.05}
\nabla _{a}P^{a}(q)=2E^{a}_{b}\nabla _{a}q^{b}
\end{align}
in arriving at the above expression we have used two results, Noether potential $J^{ab}$ being antisymmetric and $E_{ab}$ satisfying Bianchi identity. From the expression it is evident that if we enforce equation of motion for pure gravity, which amounts to: $E_{ab}=0$, the momentum becomes divergence free. Also appearance of Noether current explicitly in this expression shows intimate connection of Noether current with energy in all \LL models of gravity. 

So far the results are completely general, holding for any vector field $q^{a}$. Now we will specialize to the vector field $\xi ^{a}$ and will show that it leads to several remarkable results. First we start with momentum for $\xi ^{a}$ and its contraction with $u_{a}$ leading to (see App. \ref{Paper05:AppC:EHMomentum}):
\begin{align}\label{Paper05:LL:Momentum:Eq.06}
-u_{a}P^{a}(\xi)=D_{\alpha}\left(2N\chi ^{\alpha}\right)-2NE^{ab}u_{a}u_{b}
\end{align}
where the vector $\chi ^{a}$ is defined in \eq{Paper05:LL:Noether:Eq.04}. Now using the equation of motion i.e. $2E_{ab}=T_{ab}$ and integrating the above expression on a $t=\textrm{constant}$ surface bounded by $N=\textrm{constant}$ surface we get the following expression:
\begin{align}\label{Paper05:LL:Momentum:Eq.07}
\int _{\mathcal{R}}d^{D-1}x\sqrt{h}u_{a}\left\lbrace -P^{a}(\xi)+NT^{a}_{b}u^{b}\right\rbrace =\int _{\partial \mathcal{R}} d^{D-2}x~ T_{\rm loc}s
\end{align}
The expression on the left hand side represents total i.e. matter energy plus gravitational energy in a bulk region and the right hand side represents the heat content of the boundary surface. The temperature as usual is given by: $Na/2\pi$, i.e. the red-shifted Unruh-Davies temperature and $s$ is the Wald entropy density associated with the boundary surface (\eq{Paper05:SecLL11} provides the expression). The right hand side of the above expression can also be identified as half of the equipartition energy of the boundary surface. Hence \emph{the bulk energy originating from both gravity and matter is equal to the surface heat content}.

\paragraph*{Variation of Gravitational Energy}

From the above paragraph it is clear that the momentum $P^{a}$ and the corresponding gravitational energy $-u_{a}P^{a}(\xi)$ have very interesting thermodynamic properties. In this light, it seems natural to consider variation of the above momentum under various physical processes, e.g., how it changes due to processes acting on the boundary. We will first work with the arbitrary vector field $q^{a}$ and then we will specialize to the choice: $q^{a}=\xi ^{a}$. To our surprise just as in \gr\ even in \LL gravity variation of the gravitational momentum is connected to symplectic structures \cite{Julia1998,Regge1974}. 

Thus by variation of $-\sqrt{-g}P^{a}(q)$ and manipulating the terms carefully we obtain the symplectic structure as (see App. \ref{Paper05:AppC:EHMomentum})
\begin{align}\label{Paper05:LL:Momentum:Eq.08}
-\delta \left(\sqrt{-g}P^{a}\right)=\sqrt{-g}E_{pq}\delta g^{pq}q^{a}+\sqrt{-g}\omega ^{a}
+\partial _{c}\left(2\sqrt{-g}P_{p}^{~qr[a}q^{c]}\delta \Gamma ^{p}_{qr}\right)
\end{align}
where the symplectic form $\omega ^{a}$ has the following expression:
\begin{equation}\label{Paper05:LL:Momentum:Eq.09}
\sqrt{-g}\omega ^{a}\left(\delta, \pounds _{q}\right)=\delta \left(2\sqrt{-g}P_{p}^{~qra}\right)\pounds _{q}\Gamma ^{p}_{qr}
-\pounds _{q}\left(2\sqrt{-g}P_{p}^{~qra}\right)\delta \Gamma ^{p}_{qr}
\end{equation}
This expression is true for any arbitrary vector field $q^{a}$ and involves one arbitrary variation and another Lie variation along $q^{a}$.

Having obtained the general result we will now specialize to the vector field $\xi ^{a}$. Then we can use the above formalism in order to obtain the change in gravitational energy of the system due to its evolution, which is related to Lie derivative along $\xi ^{a}$. This can be achieved using a simple trick. First with the help of \eq{Paper05:LL:Momentum:Eq.09} we can compute the object $\delta [-\sqrt{h}u_{a}P^{a}(\xi)]$. The variation has the following expression (see App. \ref{Paper05:AppC:EHMomentum}):
\begin{align}\label{Paper05:LL:Momentum:Eq.10}
-\delta \left[\sqrt{h}u_{a}P^{a}(\xi)\right]+\sqrt{-g}E_{pq}\delta g^{pq}=\sqrt{h}u_{a}\omega ^{a}+\partial _{c}\left[2\sqrt{-g}h^{c}_{a}P_{p}^{~qra}\delta \Gamma ^{p}_{qr}\right]
\end{align}
This holds for an arbitrary variation, however what we are interested in is when the above variation is due to a diffeomorphism along $\xi ^{a}$. Again from App. \ref{Paper05:AppC:EHMomentum} using the field equation $2E_{ab}=T_{ab}$ we arrive at:
\begin{align}\label{Paper05:LL:Momentum:Eq.11}
\pounds _{\xi}\mathcal{H}_{grav}=\pounds _{\xi}\left(-\int _{\mathcal{R}}d^{D-1}x\sqrt{h}u_{a}P^{a}(\xi)\right)=\int _{\partial \mathcal{R}}d^{D-2}x\sqrt{q}Nr_{a}\left(T^{cd}\xi _{d}+2P_{p}^{~qra}\pounds _{\xi} \Gamma ^{p}_{qr}\right)
\end{align}
where $r_{a}$ is the unit normal to $N=\textrm{constant}$ surface within the $t=\textrm{constant}$ surface. Hence change of energy in the bulk is directly related to boundary effects (with the assumption that $T^{t}_{c}\xi ^{c}=0$). Among the two terms present on the right hand side, the first term is due to flow of matter energy across the boundary and the second term is related to our old friend $P_{p}^{~qra}\pounds _{\xi} \Gamma ^{p}_{qr}$. For pure gravity, i.e., $T_{ab}=0$, the above result takes a much simpler form as:
\begin{align}\label{Paper05:LL:Momentum:Eq.12}
\pounds _{\xi}\mathcal{H}_{grav}=\int _{\partial \mathcal{R}}d^{D-2}x\sqrt{q}Nr_{a}2P_{p}^{~qra}\pounds _{\xi} \Gamma ^{p}_{qr}
\end{align}
This can have direct influence on gravity wave propagation, i.e., the energy change in a bulk region due to gravity waves is related to surface processes and hence to $P_{p}^{~qra}\pounds _{\xi} \Gamma ^{p}_{qr}$.

Since the gravitational momentum is intimately connected to Noether current, we can use the above results for variation of gravitational momentum in order to obtain a variation of Noether current as well. From App. \ref{Paper05:AppC:EHMomentum} on imposing on-shell condition i.e. $E_{ab}=0$, we get:
\begin{align}\label{Paper05:LL:Momentum:Eq.13}
\sqrt{h}u_{a}\omega ^{a}\left(\delta ,\pounds _{q}\right)+2\delta \left(\sqrt{h}u_{a}E^{ab}q_{b}\right)&=\partial _{b}\left\lbrace \delta \left[\sqrt{h}u_{a}J^{ab}(q)\right]-2\sqrt{h}u_{a}P_{p}^{~qr[a}q^{b]}\delta \Gamma ^{p}_{qr}\right\rbrace 
\end{align}
This expression can be related to the usual Hamiltonian formulation where one relates bulk integral of $\delta (HN+H_{\alpha}N^{\alpha})$ (with $H$ and $H^{\alpha}$ correspond to constraints in the gravity theory) to a bulk and a surface contribution as: 
\begin{align}\label{Paper05:LL:Momentum:Eq.14}
\delta \int _{\mathcal{R}}d^{D-1}x \sqrt{h}(HN+H_{\alpha}N^{\alpha})=\int _{\mathcal{R}} d^{D-1}x \delta B +\int _{\partial \mathcal{R}}d^{D-2}x\delta S 
\end{align}
The variation present on the left hand side is equivalent to $\delta (2\sqrt{h}u_{a}E^{ab}\zeta _{b})$, where $\zeta ^{a}=Nu^{a}+N^{a}$. Then we can identify the bulk term on the right hand side with the symplectic current $\omega ^{a}$. Thus finally the remaining surface contribution turns out to be:
\begin{align}\label{Paper05:LL:Momentum:Eq.15}
\int _{\partial \mathcal{R}}d^{D-2}x\delta S =\int _{\partial \mathcal{R}}d^{D-2}x r_{b}\left\lbrace \delta \left[\sqrt{h}u_{a}J^{ab}(\zeta)\right]-2\sqrt{h}\left(NP_{p}^{~qra}+P_{p}^{~qr[a}N^{b]}\right)\delta \Gamma ^{p}_{qr}\right\rbrace 
\end{align}
This provides an elegant and simple interpretation of the surface term appearing in the variation of the gravitational action. Thus for \LL gravity the Hamiltonian formulation as well can have thermodynamic counterpart.

The above approach of relating surface quantities with bulk energy has also been studied earlier, notably in the context of Virasoro algebra and its associated central charge \cite{Carlip1999}. In \cite{Padmanabhan2012b} the above approach has been used to derive Wald entropy in \LL theories of gravity, which subsequently has been generalized in \cite{Chakraborty2014c} in order to calculate correction to horizon entropy in higher derivative gravity theories. It has been shown in this context that surface contribution alone is what will lead to the central charge and hence to horizon entropy. Our result strengthens the above feature by showing a connection between total energy and boundary heat energy in all \LL theories of gravity.

\subsection{Noether Current and Gravitational Dynamics from Related Lagrangians in \LL Gravity}
\label{Paper05:LL:QuadN}
So far we have been working extensively with the \LL Lagrangian and the vector field $\xi ^{a}$. In this section we will generalize the idea for a different class of Lagrangians, namely, the quadratic Lagrangian and the Surface Lagrangian. We will first demonstrate the Noether currents associated with these Lagrangians and then consider the gravitational momentum and its variation connected to the quadratic Lagrangian. 

\paragraph*{Noether Current}

We start our discussion by considering the Noether current associated with $L_{\rm quad}$, which is quadratic in $\Gamma ^{2}$ and hence is not a tensor density. However even if it is not a tensor density the variation will turn out to be a tensor density and there will be a conserved current associated with $L_{\rm quad}$. The above calculation can be simplified significantly by writing $\sqrt{-g}L_{\rm quad}=\sqrt{-g}L-L_{\rm sur}$. The variation can be evaluated by carefully defining the Lie derivative of non-tensorial objects, which is discussed in detail in App. \ref{Paper05:AppC:Quadratic}. From which we obtain the following result:
\begin{align}\label{Paper05:LL:Quad:Eq.01}
\pounds _{q}\left(\sqrt{-g}L_{\rm quad}\right)=\partial _{a}\left(\sqrt{-g}L_{\rm quad}q^{a}\right)-\partial _{a}\left(\sqrt{-g}K^{a}\right)
\end{align}
where $K^{a}$ is a non-tensorial object with the following expression
\begin{align}\label{Paper05:LL:Quad:Eq.02}
K^{a}=-2Q_{p}^{~qra}\partial _{q}\partial _{r}q^{p}
\end{align}
which in the \gr\ limit coincides with the expression provided in \cite{Padmanabhan2013a}. Then calculating the Lie derivative by treating $\sqrt{-g}L_{\rm quad}$ as a functional of the metric we arrive at a conservation law of the form: $\partial _{a}(\sqrt{-g}J^{a}_{\rm quad})=0$. From App. \ref{Paper05:AppC:Quadratic} the Noether current corresponding to $\sqrt{-g}L_{\rm quad}$ turns out to be:
\begin{align}\label{Paper05:LL:Quad:Eq.03}
\sqrt{-g}J^{a}_{\rm quad}&=2\sqrt{-g}E^{a}_{b}q^{b}+\sqrt{-g}L_{\rm quad}q^{a}-\sqrt{-g}K^{a}
\nonumber
\\
&-\pounds _{q}\left(2\sqrt{-g}Q_{p}^{~qra}\right)\Gamma ^{p}_{qr}+2(m-1)\sqrt{-g}Q_{p}^{~qra}\pounds _{q} \Gamma ^{p}_{qr}
\end{align}
which for $m=1$, i.e., for \gr\ takes the following form:
\begin{align}\label{Paper05:LL:Quad:Eq.04}
\sqrt{-g}J^{a}_{\rm quad}&=2\sqrt{-g}E^{a}_{b}q^{b}+\sqrt{-g}L_{\rm quad}q^{a}-\sqrt{-g}K^{a}-\pounds _{q}\left(2\sqrt{-g}Q_{p}^{~qra}\right)\Gamma ^{p}_{qr}
\nonumber
\\
&=2\sqrt{-g}E^{a}_{b}q^{b}+\sqrt{-g}L_{\rm quad}q^{a}-\sqrt{-g}K^{a}-N^{a}_{lm}\pounds _{q}f^{lm}
\end{align}
This exactly matches with the corresponding expression for \gr\ derived in \cite{Padmanabhan2013a}. Having obtained the Noether current for quadratic Lagrangian, we can use the expression of Noether current for \LL in order to obtain a relation between them. As shown in App. \ref{Paper05:AppC:Quadratic} this relation exactly mimics the corresponding one for \gr\ and leads to:
\begin{align}\label{Paper05:LL:Quad:Eq.05}
J^{a}_{\rm quad}=J^{a}+\nabla _{c}\left(q^{c}V^{a}-q^{a}V^{c}\right)=J^{a}+\nabla _{c}\left(2\Gamma ^{p}_{qr}Q_{p}^{~qr[c}q^{a]}\right)
\end{align}
where in arriving at the second line we have used the relation: $V^{a}=-2\Gamma ^{p}_{qr}Q_{p}^{~qra}$. Then we can introduce a Noether current associated with the surface term as well through the relation: $J^{a}=J^{a}_{\rm quad}+J^{a}_{\rm sur}$. Then the Noether current $J^{a}_{\rm sur}$ associated with $L_{\rm sur}$ has the following expression:
\begin{align}\label{Paper05:LL:Quad:Eq.06}
J^{a}_{\rm sur}=-\nabla _{c}\left(q^{c}V^{a}-q^{a}V^{c}\right)=\nabla _{b}\left(V^{[b}q^{a]}\right)=\nabla _{b}J^{ab}_{\rm sur};\qquad J^{ab}_{\rm sur}=q^{a}V^{b}-q^{b}V^{a}
\end{align}
Finally, variation of the Noether current corresponding to quadratic Lagrangian leads to (see App. \ref{Paper05:AppC:Quadratic}) 
\begin{align}\label{Paper05:LL:Quad:Eq.11}
\delta \left(\sqrt{-g}J^{a}_{\rm quad}\right)=\sqrt{-g}\omega ^{a}-\partial _{b}\left[\Gamma ^{p}_{qr}\delta\left(2\sqrt{-g}Q_{p}^{~qr[a}\right)q^{b]}\right]
+(m-1)\partial _{b}\left[\left(2\sqrt{-g}Q_{p}^{~qr[a}\right)q^{b]}
\delta \Gamma ^{p}_{qr}\right]+E^{a}
\end{align}
where the equation of motion terms $E^{a}$ can be given by
\begin{equation}\label{Paper05:LL:Quad:Eq.12}
E^{a}=\sqrt{-g}E_{pq}\delta g^{pq}q^{a}+2q^{b}\delta \left( \sqrt{-g}E^{a}_{b}\right)
\end{equation}
Thus we have derived the Noether current associated with the surface term in the \LL Lagrangian. Now if we use techniques of near horizon symmetry and hence Virasoro algebra to obtain the central charge it will lead to the correct Wald entropy. Hence, as we have emphasized earlier, the surface term alone is capable to produce the Wald entropy following Virasoro algebra and related central charge technique.   

\paragraph*{Quadratic Hamiltonian and its Variation}

Just as we have derived the gravitational momentum corresponding to the \LL Lagrangian we can use the quadratic Lagrangian $L_{\rm quad}$ to define another momentum as well. This momentum defined using quadratic Lagrangian has the following expression in \gr:
\begin{align}\label{Paper05:LL:Quad:Eq.07}
P^{a}_{\rm quad}=N^{a}_{lm}\pounds _{q}f^{lm}-\sqrt{-g}L_{\rm quad}q^{a}
\end{align} 
Just as we have done in the context of \EH action in this case as well we will write $N^{a}_{lm}\pounds _{q}f^{lm}=2\Gamma ^{p}_{qr}\pounds _{q}\left(\sqrt{-g}Q_{p}^{~qra}\right)$. Since $\Gamma ^{p}_{qr}$ involves 40 independent components while $Q_{p}^{~qra}$ has only metric degrees of freedom it is natural to set $\Gamma ^{p}_{qr}$ as momenta and $Q_{p}^{~qra}$ as coordinate. Then on being generalized to \LL gravity we obtain
\begin{align}\label{Paper05:LL:Quad:Eq.08}
P^{a}_{\rm quad}=\Gamma ^{p}_{qr}\pounds _{q}\left(2\sqrt{-g}P_{p}^{~qra}\right)-q^{a}\sqrt{-g}L_{\rm quad}
\end{align}
This object is less important than $P^{a}$, which we have discussed earlier, since this is a non-covariant object. This non-covariance is due to presence of $\Gamma ^{a}_{bc}$ and $L_{\rm quad}$ in its expression. In spite of these non-attractive features we will discuss properties of this momentum as well for completeness.  Following App. \ref{Paper05:AppC:Quadratic} we can write $P^{a}_{\rm quad}$ explicitly in terms of $P^{a}$ leading to
\begin{align}\label{Paper05:LL:Quad:Eq.09}
P^{a}_{\rm quad}=-\sqrt{-g}P^{a}-m\left[\partial _{c}\left(q^{[c}\mathcal{V}^{a]}\right)
+\sqrt{-g}K^{a}\right]-(m-1)q^{a}\partial _{c}\mathcal{V}^{c}
\end{align}
where $K^{a}$ is defined through \eq{Paper05:LL:Quad:Eq.02}. Then variation leads to the following on-shell result (see App. \ref{Paper05:AppC:Quadratic})
\begin{align}\label{Paper05:LL:Quad:Eq.10}
\delta P^{a}_{\rm quad}&=-\sqrt{-g}\omega ^{a}-m\delta \left(\sqrt{-g}K^{a}\right)
+m\partial _{c}\left[\delta \left(2\sqrt{-g}Q_{p}^{~qr[a}\right)q^{c]}\Gamma ^{p}_{qr}\right]
\nonumber
\\
&+(m-1)q^{a}\partial _{c}\left[\delta \left(2\sqrt{-g}Q_{p}^{~qrc}\right)\Gamma ^{p}_{qr}
+\left(2\sqrt{-g}Q_{p}^{~qrc}\right)\delta \Gamma ^{p}_{qr}\right]
\end{align}
In this case as well for $m=1$ we obtain the result derived for \gr\ and matches exactly with the one obtained in \cite{Padmanabhan2013a}. The term $\delta (\sqrt{-g}K^{a})$ shows explicitly the non-tensorial character of $\delta P^{a}_{\rm quad}$. This term can be eliminated by requiring a background subtraction. We have illustrated all these results for the sake of completeness. We will not pursue this further due to non-covariance nature of the results which leads to additional complications. 

\subsection{Heat Density of the Null Surfaces}\label{Paper05:LL:Sec.Null}

In this final section we will discuss heat density associated with a null surface in \LL gravity. Any null surface will be defined using a congruence of null vector $\ell ^{a}$, which are tangent as well as normal to the null surface. We will also assume that the null vector $\ell ^{a}$ satisfies the relation $\ell ^{2}=0$, everywhere. The null congruence will be taken to be non-affinely parametrized, such that, $\ell _{a}=A(x)\nabla _{a}B(x)$. Then the non-affinity parameter $\kappa$ can be obtained from the relation:
\begin{align}\label{Paper05:LL:Null:Eq.01}
\ell ^{j}\nabla _{j}\ell ^{i}=\kappa \ell ^{i};\qquad \kappa =\ell ^{i}\partial _{i}\ln A
\end{align}
Since the null vector is both tangent and orthogonal in order to define a projection we need an auxiliary null vector $k^{a}$ \cite{Parattu2015a} with the following properties $k^{2}=0$ and $k_{a}\ell ^{a}=-1$. Then we can introduce a projection tensor $q^{a}_{b}=\delta ^{a}_{b}+\ell ^{a}k_{b}+\ell _{b}k^{a}$ and define an associated covariant derivative $D_{a}$. 

In the previous sections we have calculated Noether current for $\xi ^{a}$ and its contraction with $u_{a}$ yielding Noether charge. In the context of null surfaces we will discuss Noether current for $\ell ^{a}$ and its contraction with $\ell _{a}$ itself. This leads to (see App. \ref{Paper05:AppC:Null})
\begin{align}\label{Paper05:LL:Null:Eq.02}
\ell _{a}J^{a}(\ell)=\nabla _{a}\left(\mathcal{K}\ell ^{a}\right)-\kappa \mathcal{K}
\end{align}
where we have introduced the object $\mathcal{K}=-2P^{abcd}\ell _{a}k_{b}\ell _{d}\nabla _{c}\ln A$. This in the \gr\ limit coincides with $\kappa$. Note the formal similarity of the expression on the right hand side, i.e., $\nabla _{a}\left(\mathcal{K}\ell ^{a}\right)-\kappa \mathcal{K}$ for null surface to the expression $D_{\alpha}\chi ^{\alpha}=\nabla _{i}\chi ^{i}-a_{j}\chi ^{j}$ obtained for spacelike surface. Also the expression for $\chi ^{a}$ is quiet similar to the expression for $\mathcal{K}$ with $\ell _{a}$ and $k_{a}$ identified with $u_{a}$ and $r_{a}$ respectively. Then introducing the covariant derivative $D_{a}$ on the surface the above expression can be written as (see App. \ref{Paper05:AppC:Null})
\begin{align}\label{Paper05:LL:Null:Eq.03}
\ell _{a}J^{a}(\ell)=2\mathcal{R}_{ab}\ell ^{a}\ell ^{b}+2\ell _{a}P_{p}^{~qra}\pounds _{\ell}\Gamma ^{p}_{qr}=D_{a}\left(\mathcal{K}\ell ^{a}\right)+\dfrac{d\mathcal{K}}{d\lambda}
\end{align}
Integrating the above expression over the null surface with integration measure $d\lambda d^{D-2}x\sqrt{q}$ and ignoring the boundary contribution we arrive at:
\begin{align}\label{Paper05:LL:Null:Eq.04}
\int d\lambda d^{D-2}x \sqrt{q}\ell _{a}J^{a}(\ell)=\int d\lambda d^{D-2}x \sqrt{q}\dfrac{d\mathcal{K}}{d\lambda}
\end{align}
This result at the face value shows that for the null surface contraction of the Noether current along $\ell _{a}$ is related to \emph{heating} of the boundary surface, with $\mathcal{K}$ being taken as temperature in the \LL gravity. 

This result is also important from the point of view of variational principle for null surfaces. Such a variational principle based on null surfaces has been carefully investigated in \cite{Padmanabhan2007}. There it was shown that given a null surface with null congruence $\ell _{a}$ we can construct the functional
\begin{align}\label{Paper05:LL:Null:Eq.05}
\mathcal{Q}=\int _{\lambda _{1}}^{\lambda _{2}}d\lambda d^{D-2}x\sqrt{q}\left(-2\mathcal{R}_{ab}+T_{ab}\right)\ell ^{a}\ell ^{b}
\end{align}
Then varying the above action functional with respect to \emph{all} $\ell _{a}$ with the constraint $\ell ^{2}=0$, we will arrive at: $\mathcal{R}^{a}_{b}-(1/2)T^{a}_{b}=f(x)\delta ^{a}_{b}$. This on using Bianchi identity, $\nabla _{a}E^{ab}=0=\nabla _{a}T^{ab}$ leads to Einstein's equation with an undetermined cosmological constant originating from integration constant. 

Then from \eq{Paper05:LL:Null:Eq.03} we can write $\mathcal{R}_{ab}\ell ^{a}\ell ^{b}$ in terms of Lie derivative of the connection and change of $\mathcal{K}$ along the null geodesics such that
\begin{align}\label{Paper05:LL:Null:Eq.06}
-2\mathcal{R}_{ab}\ell ^{a}\ell ^{b}=2\ell _{a}P_{p}^{~qra}\pounds _{\ell}\Gamma ^{p}_{qr}-\left[D_{a}\left(\mathcal{K}\ell ^{a}\right)+\dfrac{d\mathcal{K}}{d\lambda}\right]
\end{align}
Then substitution of $-2\mathcal{R}_{ab}\ell ^{a}\ell ^{b}$ on the right hand side of \eq{Paper05:LL:Null:Eq.05} leads to the following modified variational principle for \LL gravity when boundary terms are neglected as
\begin{align}\label{Paper05:LL:Null:Eq.07}
\mathcal{Q}=\int _{\lambda _{1}}^{\lambda _{2}}d\lambda d^{D-2}x\sqrt{q}\left[\left(2\ell _{a}P_{p}^{~qra}\pounds _{\ell}\Gamma ^{p}_{qr}-\dfrac{d\mathcal{K}}{d\lambda}\right)+T_{ab}\ell ^{a}\ell ^{b} \right]
\end{align}
Hence varying this Lagrangian with respect to $\ell ^{a}$ with $\ell ^{2}=0$, we can obtain the field equation for gravity with an arbitrary cosmological constant. In the above expression $T_{ab}\ell ^{a}\ell ^{b}$ can be taken as matter heat density $Ts$, while the rest of the terms represent \emph{heat density} of the null surface itself. 

We can always choose the parameter $\lambda$ such that the null vector $\ell _{a}$ is affinely parametrized. In which case $\mathcal{K}=0$ and the variational principle can be based on the following integral:
\begin{align}\label{Paper05:LL:Null:Eq.08}
\bar{\mathcal{Q}}=\int _{\lambda _{1}}^{\lambda _{2}}d\lambda d^{D-2}x\sqrt{q}\left[2\ell _{a}P_{p}^{~qra}\pounds _{\ell}\Gamma ^{p}_{qr}+T_{ab}\ell ^{a}\ell ^{b} \right]
\end{align}
Again showing explicitly the importance of the Lie derivative term in the derivation of the field equation from an alternative action principle. When there is no matter present the variational principle simplifies considerably leading to
\begin{align}\label{Paper05:LL:Null:Eq.09}
\bar{\mathcal{Q}}=\int _{\lambda _{1}}^{\lambda _{2}}d\lambda d^{D-2}x\sqrt{q}\left[2\ell _{a}P_{p}^{~qra}\pounds _{\ell}\Gamma ^{p}_{qr}\right]
\end{align}
This leads to vacuum field equation when varied over all null surfaces simultaneously. Also integral of this object has an interpretation of heat content over the boundary surface. Thus we observe that at least for affinely parametrized null congruences the variational principle over the null surface acquire a thermodynamic interpretation.

\section{Discussion}

In this work our aim was to generalize various results derived in the context of \gr\ to all \LL theories of gravity. This is a \emph{non-trivial} task since the \LL Lagrangian contains higher order terms constructed out of the curvature tensor. Also validity of some result in \gr\ does not guarantee its validity in these higher order theories, e.g., the expression for entropy in \gr\ does not hold in \LL gravity. Thus this exercise is extremely important since the \LL Lagrangian encompasses a great variety of Lagrangians all of them yielding second order equation of motion. 

Let us now summarize the key results obtained through this exercise:
\begin{itemize}
\item
In \sect{Paper05:Sec:02} we have shown that the field equation for \LL gravity near an arbitrary null surface is equivalent to a thermodynamic identity. Using a parametrization (known as GNC) for the arbitrary null surface we have shown that the field equation for \LL gravity can be used to relate energy momentum tensor with thermodynamic features. The energy-momentum tensor appears in a particular combination and can be related to the work function. Temperature related to the null surface can be obtained by invoking null geodesics and entropy can be given by the Wald entropy. This exercise also provides us a definition of energy in an arbitrary spacetime in \LL gravity, which in the static case and spherically symmetric case reduces to standard energy definitions. This energy expression in general involves $(D-2)$ dimensional \LL Lagrangian, time derivatives of the $(D-2)$-metric on the null surface. This generalizes all the previous works and demonstrates that the field equation in all \LL 
theories of gravity can be interpreted as a thermodynamic identity.
\item
In \sect{Paper05:Sec:LL:Noether} we have shown that the evolution of the spacetime in \LL gravity can be described nicely in terms of the difference $N_{\rm sur}-N_{\rm bulk}$. Here $N_{\rm sur}$ represents suitably defined surface degrees of freedom and $N_{\rm bulk}$ represents the bulk degrees of freedom. For static spacetimes, which can be thought of as equilibrium configurations we have $N_{\rm sur}=N_{\rm bulk}$, i.e., holographic equipartition holds. Thus for static spacetime degrees of freedom in the surface and in the bulk are identical. While departure from this equality is what drives the spacetime evolution in \LL gravity.
\item
Next, in \sect{Paper05:Sec:LL:Bulk} we have introduced a gravitational momentum starting from its expression in \gr. We have shown that it is intimately connected to Noether current and total gravitational plus matter energy in a bulk volume equals the heat density associated with the boundary surface. Also variation of the gravitational Hamiltonian is directly related to a symplectic structure such that time evolution of this gravitational Hamiltonian in a bulk region can be related to surface effects especially with $P_{p}^{~qra}\pounds _{\xi}\Gamma ^{p}_{qr}$. Also using this formalism it is possible to connect standard Hamiltonian formalism with the thermodynamic features discussed here.
\item
In the subsequent section \ref{Paper05:LL:QuadN} we have discussed Lie variation of non-tensorial objects and hence Noether currents associated with the quadratic and surface Lagrangians respectively. We have also introduced another momentum connected to the quadratic Lagrangian and its variation. However due to non-tensorial nature of this object, we have not pursued it further. 
\item
Finally, in \sect{Paper05:LL:Sec.Null} we have discussed an alternative variational principle for the null surfaces in \LL gravity. It turns out that the variational principle has a nice separation into matter heat density and gravitational heat density associated with the null surface. In this case as well for affine parametrization of null vectors and in vacuum spacetime, i.e., with no matter, the action functional is simply $2\sqrt{-g}P_{p}^{~qra}\pounds _{\xi}\Gamma ^{p}_{qr}$. This provides yet again thermodynamic interpretation for this Lie variation term.  
\end{itemize}

All these results suggest the importance of Noether current in any \LL theories of gravity and its relation to the thermodynamic features.  

\section*{Acknowledgement}

The author is funded by a SPM fellowship from CSIR, Government of India. The author thanks Prof. T. Padmanabhan for suggesting this problem and for providing valuable comments during preparation of the manuscript. The author also thanks Krishnamohan Parattu, Kinjalk Lochan, Suprit Singh for helpful discussions and the referee for useful structural comments about the manuscript.
\appendix

\section{Appendix: Detailed Derivations}\label{Paper05:App}

\subsection{Detailed Expressions Regarding First Law}\label{Paper05:AppA}

Let us start with evaluating the following expression in GNC coordinates introduced in the main text, which leads to
\begin{align}\label{Paper05:AppA:01}
\frac{1}{2}\left(E^{u}_{u}+E^{r}_{r}\right)&=E^{u}_{u}=E^{r}_{r}
\nonumber
\\
&=-\frac{1}{2}\frac{1}{16\pi}\frac{1}{2^{m}}\delta ^{ra_{1}b_{1}\ldots a_{m}b_{m}}_{rc_{1}d_{1}\ldots c_{m}d_{m}}R^{c_{1}d_{1}}_{a_{1}b_{1}}\ldots R^{c_{m}d_{m}}_{a_{m}b_{m}}
\nonumber
\\
&=-\frac{m}{16\pi}\frac{1}{2^{m-1}}\delta ^{ruPA_{1}B_{1}\ldots A_{m-1}B_{m-1}}_{ruQC_{1}D_{1}\ldots C_{m-1}D_{m-1}}R^{uQ}_{uP}R^{C_{1}D_{1}}_{A_{1}B_{1}}\ldots R^{C_{m-1}D_{m-1}}_{A_{m-1}B_{m-1}}
\nonumber
\\
&-\frac{m(m-1)}{16\pi}\frac{1}{2^{m-1}}\delta ^{ruPA_{1}B_{1}\ldots A_{m-1}B_{m-1}}_{ruQC_{1}D_{1}\ldots C_{m-1}D_{m-1}}R^{QC_{1}}_{uP}R^{uD_{1}}_{A_{1}B_{1}}\ldots R^{C_{m-1}D_{m-1}}_{A_{m-1}B_{m-1}}
\nonumber
\\
&-\frac{1}{16\pi}\frac{1}{2^{m+1}}\delta ^{rA_{1}B_{1}\ldots A_{m}B_{m}}_{rC_{1}D_{1}\ldots C_{m}D_{m}}R^{C_{1}D_{1}}_{A_{1}B_{1}}\ldots R^{C_{m}D_{m}}_{A_{m}B_{m}}
\end{align}
Then we obtain:
\begin{align}\label{Paper05:AppA:02}
T^{r}_{r}&=2E^{r}_{r}=-\frac{m}{8\pi}\frac{1}{2^{m-1}}\delta ^{PA_{1}B_{1}\ldots A_{m-1}B_{m-1}}_{QC_{1}D_{1}\ldots C_{m-1}D_{m-1}}R^{uQ}_{uP}R^{C_{1}D_{1}}_{A_{1}B_{1}}\ldots R^{C_{m-1}D_{m-1}}_{A_{m-1}B_{m-1}}
\nonumber
\\
&-\frac{m(m-1)}{8\pi}\frac{1}{2^{m-1}}\delta ^{PA_{1}B_{1}\ldots A_{m-1}B_{m-1}}_{QC_{1}D_{1}\ldots C_{m-1}D_{m-1}}R^{QC_{1}}_{uP}R^{uD_{1}}_{A_{1}B_{1}}\ldots R^{C_{m-1}D_{m-1}}_{A_{m-1}B_{m-1}}
\nonumber
\\
&-\frac{1}{16\pi}\frac{1}{2^{m}}\delta ^{A_{1}B_{1}\ldots A_{m}B_{m}}_{C_{1}D_{1}\ldots C_{m}D_{m}}R^{C_{1}D_{1}}_{A_{1}B_{1}}\ldots R^{C_{m}D_{m}}_{A_{m}B_{m}}
\end{align}
Now we have the following expression for components of Riemann tensor as:
\begin{subequations}
\begin{align}
R^{uQ}_{uP}&=-\frac{1}{2}q^{QE}\partial _{u}\partial _{r}q_{PE}-\frac{1}{2}q^{QE}\partial _{P}\beta _{E}-\frac{1}{2}\alpha q^{QE}\partial _{r}q_{PE}-\frac{1}{4}\beta ^{Q}\beta _{P}
\nonumber
\\
&+\frac{1}{4}\left(q^{QE}\partial _{r}q_{PF}\right)\left(q^{FL}\partial _{u}q_{EL}\right)+\frac{1}{2}q^{QE}\beta _{A}\hat{\Gamma}^{A}_{EP}
\label{Paper05:AppA:03a}
\\
R^{AM}_{CD}&=\hat{R}^{AM}_{CD}-\frac{1}{4}q^{AE}q^{MB}\Big \lbrace \partial _{u}q_{CE}\partial _{r}q_{BD}+\partial _{r}q_{CE}\partial _{u}q_{BD}-\left(C\leftrightarrow D\right)\Big \rbrace
\label{Paper05:AppA:03b}
\\
R^{uN}_{CD}&=-\frac{1}{2}q^{MN}\partial _{r}\partial _{C}q_{MD}-\frac{1}{4}\beta _{C}q^{MN}\partial _{r}q_{MD}-\frac{1}{2}\left(q^{MN}\partial _{r}q_{CE}\right)\hat{\Gamma} ^{E}_{~MD}-\left(C\leftrightarrow D\right)
\label{Paper05:AppA:03c}
\\
R^{AB}_{uC}&=q^{BD}\Big[\partial _{u}\hat{\Gamma} ^{A}_{~DC}-\frac{1}{2}q^{AE}\partial _{C}\partial _{u}q_{DE}-\frac{1}{2}\partial _{C}q^{AE}\partial _{u}q_{DE}+\frac{1}{4}\beta ^{A}\partial _{u}q_{CD}
\nonumber
\\
&+\frac{1}{2}q^{AF}\partial _{u}q_{EF}\hat{\Gamma} ^{E}_{~CD}-\frac{1}{4}\beta _{D}q^{AE}\partial _{u}q_{EC}-\frac{1}{2}q^{EF}\partial _{u}q_{FD}\hat{\Gamma} ^{A}_{~CE}\Big]
\label{Paper05:AppA:03d}
\end{align}
\end{subequations}
where $\hat{A}$ denotes an object $A$ constructed solely from the transverse metric $q_{AB}$. Note that for $\partial _{u}g_{AB}=0$, we have:
\begin{subequations}
\begin{align}
R^{uQ}_{uP}&=-\frac{1}{2}q^{QE}\partial _{P}\beta _{E}-\frac{1}{2}\alpha q^{QE}\partial _{r}q_{PE}-\frac{1}{4}\beta ^{Q}\beta _{P}+\frac{1}{2}q^{QE}\beta _{A}\hat{\Gamma}^{A}_{EP}
\label{Paper05:AppA:04a}
\\
R^{AM}_{CD}&=\hat{R}^{AM}_{CD}
\label{Paper05:AppA:04b}
\\
R^{uN}_{CD}&=-\frac{1}{2}q^{MN}\partial _{r}\partial _{C}q_{MD}-\frac{1}{4}\beta _{C}q^{MN}\partial _{r}q_{MD}-\frac{1}{2}\left(q^{MN}\partial _{r}q_{CE}\right)\hat{\Gamma} ^{E}_{~MD}-\left(C\leftrightarrow D\right)
\label{Paper05:AppA:04c}
\\
R^{AB}_{uC}&=0
\label{Paper05:AppA:04d}
\end{align}
\end{subequations}
Thus we finally arrive at the following expression:
\begin{align}\label{Paper05:AppA:05}
T^{r}_{r}&=-\frac{m}{8\pi}\frac{1}{2^{m-1}}\delta ^{PA_{1}B_{1}\ldots A_{m-1}B_{m-1}}_{QC_{1}D_{1}\ldots C_{m-1}D_{m-1}}\left(-\frac{1}{2}\alpha q^{QE}\partial _{r}q_{PE}\right)R^{C_{1}D_{1}}_{A_{1}B_{1}}\ldots R^{C_{m-1}D_{m-1}}_{A_{m-1}B_{m-1}}
\nonumber
\\
&-\frac{m}{8\pi}\frac{1}{2^{m-1}}\delta ^{PA_{1}B_{1}\ldots A_{m-1}B_{m-1}}_{QC_{1}D_{1}\ldots C_{m-1}D_{m-1}}\Big[-\frac{1}{2}q^{QE}\partial _{u}\partial _{r}q_{PE}-\frac{1}{2}q^{QE}\partial _{P}\beta _{E}-\frac{1}{4}\beta ^{Q}\beta _{P}
\nonumber
\\
&+\frac{1}{4}\left(q^{QE}\partial _{r}q_{PF}\right)\left(q^{FL}\partial _{u}q_{EL}\right)+\frac{1}{2}q^{QE}\beta _{A}\hat{\Gamma}^{A}_{EP}\Big]R^{C_{1}D_{1}}_{A_{1}B_{1}}\ldots R^{C_{m-1}D_{m-1}}_{A_{m-1}B_{m-1}}
\nonumber
\\
&-\frac{m(m-1)}{8\pi}\frac{1}{2^{m-1}}\delta ^{PA_{1}B_{1}\ldots A_{m-1}B_{m-1}}_{QC_{1}D_{1}\ldots C_{m-1}D_{m-1}}R^{QC_{1}}_{uP}R^{uD_{1}}_{A_{1}B_{1}}\ldots R^{C_{m-1}D_{m-1}}_{A_{m-1}B_{m-1}}
\nonumber
\\
&-\frac{1}{16\pi}\frac{1}{2^{m}}\delta ^{A_{1}B_{1}\ldots A_{m}B_{m}}_{C_{1}D_{1}\ldots C_{m}D_{m}}R^{C_{1}D_{1}}_{A_{1}B_{1}}\ldots R^{C_{m}D_{m}}_{A_{m}B_{m}}
\end{align}
which can be simplified and finally leads to the following expression:
\begin{align}\label{Paper05:AppA:05new}
T^{r}_{r}&=2E^{r}_{r}=\left(E^{u}_{u}+E^{r}_{r}\right)
\nonumber
\\
&=\frac{m}{8}\frac{1}{2^{m-1}}\left(\frac{\alpha}{2\pi}\right)\left(\delta ^{PA_{1}B_{1}\ldots A_{m-1}B_{m-1}}_{QC_{1}D_{1}\ldots C_{m-1}D_{m-1}}R^{C_{1}D_{1}}_{A_{1}B_{1}}\ldots R^{C_{m-1}D_{m-1}}_{A_{m-1}B_{m-1}}\right)\left(q^{QE}\partial _{r}q_{PE}\right)
\nonumber
\\
&-\frac{m}{8\pi}\frac{1}{2^{m-1}}\delta ^{PA_{1}B_{1}\ldots A_{m-1}B_{m-1}}_{QC_{1}D_{1}\ldots C_{m-1}D_{m-1}}\Big[-\frac{1}{2}q^{QE}\partial _{u}\partial _{r}q_{PE}-\frac{1}{2}q^{QE}\partial _{P}\beta _{E}-\frac{1}{4}\beta ^{Q}\beta _{P}
\nonumber
\\
&+\frac{1}{4}\left(q^{QE}\partial _{r}q_{PF}\right)\left(q^{FL}\partial _{u}q_{EL}\right)+\frac{1}{2}q^{QE}\beta _{A}\hat{\Gamma}^{A}_{EP}\Big]R^{C_{1}D_{1}}_{A_{1}B_{1}}\ldots R^{C_{m-1}D_{m-1}}_{A_{m-1}B_{m-1}}
\nonumber
\\
&-\frac{m(m-1)}{8\pi}\frac{1}{2^{m-1}}\delta ^{PA_{1}B_{1}\ldots A_{m-1}B_{m-1}}_{QC_{1}D_{1}\ldots C_{m-1}D_{m-1}}R^{QC_{1}}_{uP}R^{uD_{1}}_{A_{1}B_{1}}\ldots R^{C_{m-1}D_{m-1}}_{A_{m-1}B_{m-1}}
\nonumber
\\
&-\frac{1}{16\pi}\frac{1}{2^{m}}\delta ^{A_{1}B_{1}\ldots A_{m}B_{m}}_{C_{1}D_{1}\ldots C_{m}D_{m}}R^{C_{1}D_{1}}_{A_{1}B_{1}}\ldots R^{C_{m}D_{m}}_{A_{m}B_{m}}
\end{align}
This is the expression used in the text. We also have entropy density to be:
\begin{align}\label{Paper05:AppA:06}
s&=4\pi m\sqrt{q}\mathcal{L}_{m-1}^{(D-2)}
\nonumber
\\
&=4\pi m \sqrt{q}\left(\frac{1}{16\pi}\frac{1}{2^{m-1}}\delta ^{A_{1}B_{1}\ldots A_{m-1}B_{m-1}}_{C_{1}D_{1}\ldots C_{m-1}D_{m-1}}R^{C_{1}D_{1}}_{A_{1}B_{1}}\ldots R^{C_{m-1}D_{m-1}}_{A_{m-1}B_{m-1}} \right)
\end{align}
Then under variation along the radial coordinate i.e. along $k^{a}$ parametrized by $\lambda$ we have:
\begin{align}\label{Paper05:AppA:07}
\delta _{\lambda}s&=4\pi m \left(\frac{1}{2}q^{AB}\delta _{\lambda}q_{AB}\right)\sqrt{q}\mathcal{L}_{m-1}^{(D-2)}
\nonumber
\\
&-4\pi m \sqrt{q}\left(\frac{m-1}{16\pi}\frac{1}{2^{m-1}}\delta ^{A_{1}B_{1}\ldots A_{m-1}B_{m-1}}_{C_{1}D_{1}\ldots C_{m-1}D_{m-1}}R^{C_{1}A}_{A_{1}B_{1}}q^{D_{1}B}\delta _{\lambda}q_{AB}\ldots R^{C_{m-1}D_{m-1}}_{A_{m-1}B_{m-1}} \right)
\nonumber
\\
&=-4\pi m \sqrt{q}\delta _{\lambda}q_{AB}\Big(-\frac{1}{2}q^{AB}\mathcal{L}_{m-1}^{(D-2)}
\nonumber
\\
&+\frac{m-1}{16\pi}\frac{1}{2^{m-1}}q^{BD_{1}}\delta ^{A_{1}B_{1}\ldots A_{m-1}B_{m-1}}_{D_{1}C_{1}\ldots C_{m-1}D_{m-1}}R^{AC_{1}}_{A_{1}B_{1}}\ldots R^{C_{m-1}D_{m-1}}_{A_{m-1}B_{m-1}}\Big)
\nonumber
\\
&=-4\pi m\sqrt{q}E^{AB}\delta _{\lambda}q_{AB}
\end{align}
where we have:
\begin{align}\label{Paper05:AppA:08}
E^{A}_{B}&=-\frac{1}{2}\delta ^{A}_{B}\mathcal{L}_{m-1}^{(D-2)}+\frac{m-1}{16\pi}\frac{1}{2^{(m-1)}}\delta ^{A_{1}B_{1}\ldots A_{m-1}B_{m-1}}_{BC_{1}\ldots C_{m-1}D_{m-1}}R^{AC_{1}}_{A_{1}B_{1}}\ldots R^{C_{m-1}D_{m-1}}_{A_{m-1}B_{m-1}}
\nonumber
\\
&=-\frac{1}{2}\frac{1}{16\pi}\frac{1}{2^{m-1}}\delta ^{AA_{1}B_{1}\ldots A_{m-1}B_{m-1}}_{BC_{1}D_{1}\ldots C_{m-1}D_{m-1}}R^{C_{1}D_{1}}_{A_{1}B_{1}}\ldots R^{C_{m-1}D_{m-1}}_{A_{m-1}B_{m-1}}
\end{align}
Hence we obtain:
\begin{align}\label{Paper05:AppA:09}
\delta _{\lambda}s&=-4\pi m\sqrt{q}\delta _{\lambda}q_{AB}\Big(-\frac{1}{2}\frac{1}{16\pi}\frac{1}{2^{m-1}}\delta ^{AA_{1}B_{1}\ldots A_{m-1}B_{m-1}}_{BC_{1}D_{1}\ldots C_{m-1}D_{m-1}}
\nonumber
\\
&\times R^{C_{1}D_{1}}_{A_{1}B_{1}}\ldots R^{C_{m-1}D_{m-1}}_{A_{m-1}B_{m-1}}\Big)
\nonumber
\\
&=\frac{m}{8 2^{m-1}}\sqrt{q}\left(\delta ^{AA_{1}B_{1}\ldots A_{m-1}B_{m-1}}_{BC_{1}D_{1}\ldots C_{m-1}D_{m-1}}
R^{C_{1}D_{1}}_{A_{1}B_{1}}\ldots R^{C_{m-1}D_{m-1}}_{A_{m-1}B_{m-1}}\right)q^{BC}\delta _{\lambda}q_{AC}
\end{align}
Finally using \eq{Paper05:AppA:09} in \eq{Paper05:AppA:05new} we obtain the most general expression for energy as
\begin{align}
\delta _{\lambda}E&=\delta \lambda\int d\Sigma \Big\lbrace \frac{m}{8\pi}\frac{1}{2^{m-1}}\delta ^{PA_{1}B_{1}\ldots A_{m-1}B_{m-1}}_{QC_{1}D_{1}\ldots C_{m-1}D_{m-1}}\Big[-\frac{1}{2}q^{QE}\partial _{u}\partial _{r}q_{PE}-\frac{1}{2}q^{QE}\partial _{P}\beta _{E}-\frac{1}{4}\beta ^{Q}\beta _{P}
\nonumber
\\
&+\frac{1}{4}\left(q^{QE}\partial _{r}q_{PF}\right)\left(q^{FL}\partial _{u}q_{EL}\right)+\frac{1}{2}q^{QE}\beta _{A}\hat{\Gamma}^{A}_{EP}\Big]R^{C_{1}D_{1}}_{A_{1}B_{1}}\ldots R^{C_{m-1}D_{m-1}}_{A_{m-1}B_{m-1}}
\nonumber
\\
&+\frac{m(m-1)}{8\pi}\frac{1}{2^{m-1}}\delta ^{PA_{1}B_{1}\ldots A_{m-1}B_{m-1}}_{QC_{1}D_{1}\ldots C_{m-1}D_{m-1}}R^{QC_{1}}_{uP}R^{uD_{1}}_{A_{1}B_{1}}\ldots R^{C_{m-1}D_{m-1}}_{A_{m-1}B_{m-1}}
\nonumber
\\
&+\frac{1}{16\pi}\frac{1}{2^{m}}\delta ^{A_{1}B_{1}\ldots A_{m}B_{m}}_{C_{1}D_{1}\ldots C_{m}D_{m}}R^{C_{1}D_{1}}_{A_{1}B_{1}}\ldots R^{C_{m}D_{m}}_{A_{m}B_{m}}\Big\rbrace
\end{align}
where $d\Sigma =d^{D-2}x\sqrt{q}$ is the integration measure on the null surface.
\subsection{GNC metric in static form}\label{Paper05:AppB}

Conversion of GNC metric to static coordinates has been performed explicitly in \cite{Chakraborty2015a}. In this appendix, we will present a short discussion of this transformation from GNC to static coordinates (for detailed discussion see \cite{Chakraborty2015a}). First converting GNC coordinates to Rindler coordinates one can identify the timelike Killing vector $\xi ^{a}$ with the following components:
\begin{align}\label{Paper05:AppB:Eq.01}
\xi ^{a}=\left(1,0,0,0\right);\qquad
\xi _{a}=\left(-2r\alpha ,1,-r\beta _{A}\right)
\end{align} 
Imposing hypersurface orthogonality on $\xi ^{a}$, i.e. $\xi _{[a}\nabla _{b}\xi _{c]}=0$ leads to two conditions that the metric elements need to satisfy everywhere:
\begin{subequations}
\begin{align}
r\beta _{A}\partial _{r}\alpha -r\alpha \partial _{r}\beta _{A}+\partial _{A}\alpha -\frac{1}{2}\partial _{u}\beta _{A}&=0
\label{Paper05:AppB:Eq.02a}
\\
r\left(\beta _{A}\partial _{r}\beta _{B}-\beta _{B}\partial _{r}\beta _{A}\right)-\left(\partial _{B}\beta _{A}-\partial _{A}\beta _{B}\right)&=0
\label{Paper05:AppB:Eq.02b}
\end{align}
\end{subequations}
Next we need to impose Killing condition, which amounts to set $\nabla _{a}\xi _{b}+\nabla _{b}\xi _{a}=0$ everywhere in the spacetime region under our consideration. This implies \cite{Chakraborty2015a}: 
\begin{align}\label{Paper05:AppB:Eq.04}
\partial _{u}\alpha &=0;\qquad \partial _{u}\beta _{A}=0;\qquad \partial _{u}q_{AB}=0;
\end{align}
If a spacetime satisfies the Killing conditions then it is called a stationary spacetime. Like in this case if the GNC parameters $\alpha$, $\beta _{A}$ and $q_{AB}$ are independent of the $u$ coordinate then the GNC parametrization would lead to a stationary spacetime. When a spacetime is stationary if we further impose hypersurface orthogonality condition, then the spacetime is called static. In this case if we use both the Killing condition and hypersurface orthogonality for $\xi ^{a}$ then from \eq{Paper05:AppB:Eq.02b} we obtain:
\begin{align}\label{Paper05:AppB:Eq.05}
\partial _{A}\alpha \vert _{r=0}=0
\end{align}
Thus imposing staticity condition on the GNC metric with the time evolution vector field $\xi ^{a}$, we arrive at zeroth low of black hole thermodynamics generalized to an arbitrary null surface in \LL gravity. This condition we will use frequently in the main text. 
\subsection{Various Identities Used in The Text Regarding \LL Gravity}\label{Paper05:AppC}

In this subsection we will collect derivation of important identities used in the text while describing the generalization to \LL gravity. We will order the derivations as in text.

\subsubsection{Alternative derivation of Noether current}\label{Paper05:AppC:Noether}

In \EH action the Noether potential derived from diffeomorphism is:
\begin{align}
16\pi J^{ab}(v)&=\nabla ^{a}v^{b}-\nabla ^{b}v^{a}
\nonumber
\\
&=\left(\delta ^{a}_{i}\delta ^{b}_{j}-\delta ^{a}_{j}\delta ^{b}_{i}\right)\nabla ^{i}v^{j}
\nonumber
\\
&=32\pi Q^{ab}_{ij}\nabla ^{i}v^{j}
\end{align}
We can use this result to motivate the Noether current for LL gravity as:
\begin{align}
J^{ab}=2P^{ab}_{cd}\nabla ^{c}v^{d}=2P^{abcd}\nabla _{c}v_{d}
\end{align}
where $P^{abcd}$ is assumed to be antisymmetric in $(a,b)$ and $(c,d)$. However to get the 
standard expressions for Noether current we also assume that $\nabla _{b}P^{abcd}=0$, 
which by antisymmetry automatically imply: $\nabla _{a}P^{abcd}=0$. Note that we have not 
assumed anything regarding cyclic identity or vanishing of divergence in the last two indices.  
Then from the relation:
\begin{align}
\pounds _{v}\Gamma ^{a}_{bc}=\nabla _{b}\nabla _{c}v^{a}+R^{a}_{~cmb}v^{m}
\end{align}
we can write the Noether current in the usual way as:
\begin{align}
J^{a}&=\nabla _{b}J^{ab}=2P^{abcd}\nabla _{b}\nabla _{c}v_{d}
\nonumber
\\
&=2P^{abcd}\left[g_{pd}\pounds _{v}\Gamma ^{p}_{bc}-R_{dcmb}v^{m}\right]
\nonumber
\\
&=-2P^{abcd}R_{dcmb}v^{m}+2g_{pd}P^{abcd}\pounds _{v}\Gamma ^{p}_{bc}
\nonumber
\\
&=2P^{abcd}R_{mbcd}v^{m}+2g_{pd}P^{abcd}\pounds _{v}\Gamma ^{p}_{bc}
\nonumber
\\
&=2\mathcal{R}^{a}_{m}v^{m}+2P_{p}^{~cba}\pounds _{v}\Gamma ^{p}_{bc}
\end{align}
In the last line we have used the fact that $P^{abcd}$ is symmetric 
in the pair exchange $(a,b)$ and $(c,d)$. Then the connection 
between Noether currents of two vectors $v^{a}=f(x)q^{a}$ and 
$q^{a}$ itself has been obtained in an earlier work and can be 
derived depending on only the above mentioned properties of the 
tensor $P^{abcd}$. 

However we will now present an alternative derivation. Let us start with the following identity:
\begin{align}
\nabla _{a}\xi _{b}&=\nabla _{a}\left(Nu_{b}\right)=N\nabla _{a}u_{b}+u_{b}\nabla _{a}N
\nonumber
\\
&=N\nabla _{a}u_{b}+u_{b}\left(Na_{a}-u_{a}u^{i}\nabla _{i}N\right)
\nonumber
\\
&=-N\left(K_{ab}+u_{a}a_{b}\right)+Na_{a}u_{b}-\left(u^{i}\nabla _{i}N\right)u_{a}u_{b}
\end{align}
Then the Noether potential corresponding to the vector field $\xi ^{a}$ 
turns out to be:
\begin{align}
J^{ab}(\xi)&=2P^{abcd}\nabla _{c}\xi _{d}
\nonumber
\\
&=2P^{abcd}\left[-N\left(K_{ab}+u_{a}a_{b}\right)+Na_{a}u_{b}
-\left(u^{i}\nabla _{i}N\right)u_{a}u_{b} \right]
\nonumber
\\
&=-2NaP^{abcd}\left(u_{c}\hat{a}_{d}-\hat{a}_{c}u_{d}\right)
\nonumber
\\
&=-2NaP^{abcd}\epsilon _{cd}
\end{align}
where $\epsilon _{cd}$ is the bi-normal to the $N=\textrm{constant}$ surface 
within $t=\textrm{constant}$ surface. The same for Noether potential of the 
four velocity vector leads to:
\begin{align}
J^{ab}(u)&=2P^{abcd}\nabla _{c}u_{d}
\nonumber
\\
&=-2P^{abcd}\left(K_{cd}+u_{c}a_{d}\right)
\nonumber
\\
&=-aP^{abcd}\epsilon _{cd}
\end{align}
These are the expressions used in the text. From the above expressions the Noether current for $\xi ^{a}$ turns out to be:
\begin{align}
u_{a}J^{a}(\xi)&=u_{a}\nabla _{b}J^{ab}(\xi)=u_{a}\nabla _{b}\left[2NJ^{ab}(u)\right]
\nonumber
\\
&=2Nu_{a}J^{a}(u)+2u_{a}\nabla _{b}NJ^{ab}(u)
\nonumber
\\
&=2Nu_{a}J^{a}(u)+2u_{a}\left(Na_{b}-u_{b}u^{i}\nabla _{i}N\right)J^{ab}(u)
\nonumber
\\
&=2Nu_{a}J^{a}(u)+2Nu_{a}a_{b}J^{ab}(u)
\end{align}
Now we have the following relations:
\begin{align}
u_{a}J^{ab}(u)&=-2P^{abcd}u_{a}u_{c}a_{d}=\chi ^{b}
\\
Nu_{a}J^{a}(u)&=Nu_{a}\nabla _{b}J^{ab}(u)
\nonumber
\\
&=\nabla _{b}\left(Nu_{a}J^{ab}(u)\right)
-J^{ab}(u)\left(u_{a}\nabla _{b}N+N\nabla _{b}u_{a}\right)
\end{align}
On using these relations we arrive at the following expression for $u_{a}J^{a}(\xi)$ as:
\begin{align}
u_{a}J^{a}(\xi)&=Nu_{a}J^{a}(u)+2Nu_{a}a_{b}J^{ab}(u)+\nabla _{b}\left(N\chi ^{b}\right)
\nonumber
\\
&-J^{ab}(u)\left[u_{a}\left(Na_{b}-u_{b}u^{i}\nabla _{i}N\right)-N\left(K_{ba}+u_{b}a_{a}\right)\right]
\nonumber
\\
&=Nu_{a}J^{a}(u)+2Nu_{a}a_{b}J^{ab}(u)+\nabla _{b}\left(N\chi ^{b}\right)
-NJ^{ab}(u)\left(u_{a}a_{b}-u_{b}a_{a}\right)
\nonumber
\\
&=Nu_{a}J^{a}(u)+\nabla _{b}\left(N\chi ^{b}\right)
\end{align}
Now we need to evaluate the quantity $u_{a}J^{a}(u)$. This quantity has the following expression:
\begin{align}
u_{a}J^{a}(u)&=u_{a}\nabla _{b}J^{ab}(u)
\nonumber
\\
&=-2P^{abcd}u_{a}\nabla _{b}\left(u_{c}a_{d}\right)
\nonumber
\\
&=\nabla _{b}\left(-2P^{abcd}u_{a}u_{c}a_{d}\right)
+2P^{abcd}u_{c}a_{d}\nabla _{b}u_{a}
\nonumber
\\
&=\nabla _{b}\left(-2P^{abcd}u_{a}u_{c}a_{d}\right)
-2P^{abcd}u_{c}a_{d}\left(K_{ba}+u_{b}a_{a}\right)
\nonumber
\\
&=\nabla _{b}\chi ^{b}-\chi ^{i}a_{i}=D_{\alpha}\chi ^{\alpha}
\end{align}
This is the result used to obtain \eq{Paper05:LL:Noether:Eq.03}. 

\subsubsection{Gravitational Momentum and related derivations for \EH Action}\label{Paper05:AppC:EHMomentum}

In this section we provide derivation to various identities used in \sect{Paper05:Sec:LL:Bulk}. We start by giving the result for variation of \LL Lagrangian:
\begin{align}
\delta \left(\sqrt{-g}L\right)=\sqrt{-g}E_{ab}\delta g^{ab}
-\partial _{c}\left(2\sqrt{-g}P_{p}^{~qrc}\delta \Gamma ^{p}_{qr}\right)
\end{align}
The Noether current can be written as:
\begin{align}
J^{a}(q)&=2\mathcal{R}^{a}_{b}q^{b}+2P_{p}^{~qra}\pounds _{q}\Gamma ^{p}_{qr}
\nonumber
\\
&=2E^{a}_{b}q^{b}+2P_{p}^{~qra}\pounds _{q}\Gamma ^{p}_{qr}+Lq^{a}
\nonumber
\\
&=2E^{a}_{b}q^{b}-P^{a}(q)
\end{align}
Then multiplying both sides by the four velocity $u_{a}$ and taking $q_{a}=\xi _{a}$ we readily obtain:
\begin{align}
-u_{a}P^{a}(\xi)&=u_{a}J^{a}(\xi)-2E^{ab}\xi _{a}u_{b}
\nonumber
\\
&=D_{\alpha}\left(2N\chi ^{\alpha}\right)-2NE^{ab}u_{a}u_{b}
\end{align}
Let us now consider the variation of the gravitational momentum corresponding to $q^{a}$. This has the following expression:
\begin{align}
-\delta \left(\sqrt{-g}P^{a}\right)&=q^{a}\delta \left(\sqrt{-g}L\right) 
+\delta \left(2\sqrt{-g}P_{p}^{~qra}\pounds _{q}\Gamma ^{p}_{qr}\right)
\nonumber
\\
&=q^{a}\left(\sqrt{-g}E_{pq}\delta g^{pq}\right)
-q^{a}\partial _{c}\left(2\sqrt{-g}P_{p}^{~qrc}\delta \Gamma ^{p}_{qr} \right)
+\pounds _{q}\left(2\sqrt{-g}P_{p}^{~qra}\delta \Gamma ^{p}_{qr} \right)
\nonumber
\\
&+\delta \left(2\sqrt{-g}P_{p}^{~qra}\right)\pounds _{q}\Gamma ^{p}_{qr}
-\pounds _{q}\left(2\sqrt{-g}P_{p}^{qra}\right)\delta \Gamma ^{p}_{qr}
\nonumber
\\
&=\sqrt{-g}E_{pq}\delta g^{pq}q^{a}+\sqrt{-g}\omega ^{a}
+\partial _{c}\left(2\sqrt{-g}P_{p}^{~qr[a}q^{c]}\delta \Gamma ^{p}_{qr}\right)
\end{align}
where in arriving at the last line we have used the following relation:
\begin{align}
\pounds _{q}Q^{a}-q^{a}\partial _{c}Q^{c}=\partial _{c}\left(q^{[c}Q^{a]}\right)
\end{align}
for the tensor density $Q^{c}=2\sqrt{-g}P_{p}^{~qrc}\delta \Gamma ^{p}_{qr}$. Also since $q^{a}$ is a constant vector $\delta$ and $\pounds _{q}$ are assumed to commute. Also the object $\omega ^{a}$ is defined as,
\begin{align}
\sqrt{-g}\omega ^{a}\left(\delta, \pounds _{q}\right)=\delta \left(2\sqrt{-g}P_{p}^{~qra}\right)\pounds _{q}\Gamma ^{p}_{qr}
-\pounds _{q}\left(2\sqrt{-g}P_{p}^{~qra}\right)\delta \Gamma ^{p}_{qr}
\end{align}
This is the result used in \eq{Paper05:LL:Momentum:Eq.09}. Then we can write the above relation in terms of the Noether current as:
\begin{align}
\delta \left(\sqrt{-g}J^{a}-2\sqrt{-g}E^{a}_{b}q_{b}\right)
=\sqrt{-g}E_{pq}\delta g^{pq}q^{a}+\sqrt{-g}\omega ^{a}
+\partial _{c}\left(2\sqrt{-g}P_{p}^{~qr[a}q^{c]}\delta \Gamma ^{p}_{qr}\right)
\end{align}
The above relation can also be written using the Noether potential as:
\begin{align}
\partial _{b}\left\lbrace \delta \left(\sqrt{-g}J^{ab}\right)-2\sqrt{-g}P_{p}^{~qr[a}q^{b]}\delta 
\Gamma ^{p}_{qr}\right\rbrace 
=\sqrt{-g}E_{pq}\delta g^{pq}q^{a}+\sqrt{-g}\omega ^{a}
+2\delta \left(\sqrt{-g}E^{ab}q_{b}\right)
\end{align}
While for on-shell (i.e., when $E_{ab}=0$) we have the following relations:
\begin{align}
-\delta \left(\sqrt{-g}P^{a}\right)&=\sqrt{-g}\omega ^{a}
+\partial _{c}\left(2\sqrt{-g}P_{p}^{~qr[a}q^{c]}\delta \Gamma ^{p}_{qr}\right)
\\
\partial _{b}\left\lbrace \delta \left(\sqrt{-g}J^{ab}\right)-2\sqrt{-g}P_{p}^{~qr[a}q^{b]}\delta 
\Gamma ^{p}_{qr}\right\rbrace 
&=\sqrt{-g}\omega ^{a}
+2\delta \left(\sqrt{-g}E^{ab}q_{b}\right)
\end{align}
Then integrating the second equation over volume $\mathcal{R}$ with $d^{D-1}x\sqrt{h}$ as integration measure and $q^{a}=\zeta ^{a}=Nu^{a}+N^{a}$ we arrive at:
\begin{align}
\delta \int _{\mathcal{R}}d^{D-1}x\sqrt{h}\left(2u_{a}E^{ab}\zeta _{b}\right)&=\int _{\mathcal{R}}d^{D-1}x\partial _{b}\Big\lbrace \delta \left[\sqrt{h}u_{a}J^{ab}(\zeta)\right]
-2\sqrt{h}u_{a}\left(NP_{p}^{~qr[a}u^{b]}+P_{p}^{~qr[a}N^{b]}\right)\delta \Gamma ^{p}_{qr}\Big\rbrace 
\nonumber
\\
&-\int _{\mathcal{R}}d^{D-1}x\sqrt{h}u_{a}\omega ^{a}\left(\delta ,\pounds _{q}\right)
\nonumber
\\
&=\int _{\mathcal{R}}d^{D-1}x\partial _{b}\Big\lbrace \delta \left[\sqrt{h}u_{a}J^{ab}(\zeta)\right]
-2\sqrt{h}\left(Nh^{b}_{a}P_{p}^{~qra}+P_{p}^{~qr[a}N^{b]}\right)\delta \Gamma ^{p}_{qr}\Big\rbrace 
\nonumber
\\
&-\int _{\mathcal{R}}d^{D-1}x\sqrt{h}u_{a}\omega ^{a}\left(\delta ,\pounds _{q}\right)
\end{align}
Then we want the variation of the Hamiltonian obtained by contracting the momentum along the four velocity $u_{a}$, such that:
\begin{align}
-\sqrt{h}u_{a}P^{a}(\xi)=t_{a}\sqrt{-g}P^{a}(\xi)
\end{align}
where the vector $t_{a}=-u_{a}/N$ in the coordinate system under consideration. Then varying the above expression (noting that variation of $t_{a}$ vanishes) we obtain
\begin{align}\label{AppC:Eq.01}
-\delta \left[\sqrt{h}u_{a}P^{a}(\xi)\right]
&=t_{a}\delta \left[\sqrt{-g}P^{a}(\xi)\right]
\nonumber
\\
&=-t_{a}\left[\sqrt{-g}E_{pq}\delta g^{pq}\xi^{a}+\sqrt{-g}\omega ^{a}
+\partial _{c}\left(2\sqrt{-g}P_{p}^{~qr[a}\xi ^{c]}\delta \Gamma ^{p}_{qr}\right)\right]
\nonumber
\\
&=\sqrt{h}u_{a}\omega ^{a}-\sqrt{-g}E_{pq}\delta g^{pq}
+\partial _{c}\left[2\sqrt{-g}u_{a}P_{p}^{~qr[a}u^{c]}\delta \Gamma ^{p}_{qr}\right]
\nonumber
\\
&=\sqrt{h}u_{a}\omega ^{a}-\sqrt{-g}E_{pq}\delta g^{pq}
+\partial _{c}\left[2\sqrt{-g}h^{c}_{a}P_{p}^{~qra}\delta \Gamma ^{p}_{qr}\right]
\end{align}
where in the second line we have used the standard trick in order to get $u_{a}$ and in the last line we have used the relation:
\begin{align}
u_{a}P_{p}^{~qr[a}u^{c]}=h^{c}_{a}P_{p}^{~qra}
\end{align}
Defining the gravitational Hamiltonian as:
\begin{equation}
\mathcal{H}_{grav}=-\int d^{D-1}x \sqrt{h}u_{a}P^{a}(\xi)
\end{equation}
its variation can be obtained readily from \eq{AppC:Eq.01} as:
\begin{equation}
\delta \mathcal{H}_{grav}=\int d^{D-1}x\sqrt{h}u_{a}\omega ^{a}-\int d^{D-1}x \sqrt{-g}E_{pq}\delta g^{pq}+\int d^{D-2}x~2r_{c}\sqrt{q}P_{p}^{~qrc}\delta \Gamma ^{p}_{qr}
\end{equation}
where in order to obtain the last term we have used the result that $r_{c}h^{c}_{a}=r_{a}$ and $\sqrt{-g}=N\sqrt{h}$. The above results are true for arbitrary variations. Applying it to Lie variation along $\xi ^{a}$ we arrive at the following form for \eq{AppC:Eq.01} as:
\begin{align}
-\pounds _{\xi}\left[\sqrt{h}u_{a}P^{a}(\xi)\right]=2\sqrt{-g}E_{pq}\nabla ^{p}\xi ^{q}+\partial _{c}\left[2\sqrt{-g}h^{c}_{a}P_{p}^{~qra}\pounds _{\xi} \Gamma ^{p}_{qr}\right]
\end{align}
In arriving at the above result we have used the relations: $\pounds _{\xi}g^{ab}=-(\nabla ^{a}\xi ^{b}+\nabla ^{b}\xi ^{a})$ and $\omega ^{a}(\pounds _{\xi},\pounds _{\xi})=0$. Now using Bianchi identity $\nabla _{a}E^{ab}=0$ we arrive at:
\begin{align}
-\pounds _{\xi}\left[\sqrt{h}u_{a}P^{a}(\xi)\right]=\partial _{c}\left[2\sqrt{-g}\left(E^{cd}\xi _{d}+h^{c}_{a}P_{p}^{~qra}\pounds _{\xi} \Gamma ^{p}_{qr}\right)\right]
\end{align}
This is the relation used to arrive at the results in the main text.

\subsubsection{Noether Current and Gravitational Momentum for Related Lagrangians}
\label{Paper05:AppC:Quadratic}

We first consider Lie derivative of various expressions which also involve non-tensorial objects. Since throughout we mostly have $\pounds _{q}\Gamma ^{a}_{bc}$ appearing in most of the expressions we will evaluate it explicitly. The Lie derivative has the following expression:
\begin{align}
\pounds _{q}\Gamma ^{a}_{bc}&=\nabla _{b}\nabla _{c}q^{a}+R^{a}_{~cmb}q^{m}
\nonumber
\\
&=q^{d}\partial _{d}\Gamma ^{a}_{bc}-\Gamma ^{d}_{bc}\partial _{d}q^{a}+q^{d}\partial _{d}\Gamma ^{a}_{bc}+\Gamma ^{a}_{dc}\partial _{b}q^{d}+q^{d}\partial _{d}\Gamma ^{a}_{bc}+\Gamma ^{a}_{bd}\partial _{c}q^{d}+\partial _{b}\partial _{c}q^{a}
\nonumber
\\
&=\left[\pounds _{q}\Gamma ^{a}_{bc}\right]_{\rm std}+\partial _{b}\partial _{c}q^{a}
\end{align} 
In this and all other expressions, the object $\left(\pounds _{q}\ldots \right)_{std}$ represents the Lie variation computed using the $\ldots$ object as tensorial. For ease of notation in the later part of the calculation we will define the following objects:
\begin{align}
\mathcal{V}^{a}&=-2\sqrt{-g}Q_{p}^{~qra}\Gamma ^{p}_{qr}
\\
V^{a}&=-2Q_{p}^{~qra}\Gamma ^{p}_{qr}=\frac{1}{\sqrt{-g}}\mathcal{V}^{a}
\end{align}
Note that both of them are non-tensorial due to presence of the connection $\Gamma ^{a}_{bc}$. Following the above procedure the Lie variation of the object $\mathcal{V}^{a}$ takes the following form:
\begin{align}
\pounds _{q}\mathcal{V}^{a}&=\pounds _{q}\left(-2\sqrt{-g}Q_{p}^{~qra}\right)\Gamma ^{p}_{qr}-2\sqrt{-g}Q_{p}^{~qra}\pounds _{q}\Gamma ^{p}_{qr}
\nonumber
\\
&=\left(\pounds _{q}\mathcal{V}^{a}\right)_{\rm std}+\sqrt{-g}K^{a}
\end{align}
where we have defined the non-tensorial part $K^{a}$ as:
\begin{equation}
K^{a}=-2Q_{p}^{~qra}\partial _{q}\partial _{r}q^{p}
\end{equation}
In the \EH limit the above non-tensorial object goes to:
\begin{equation}
16\pi K^{a}\rightarrow -\left(\delta ^{r}_{p}g^{qa}-\delta ^{a}_{p}g^{rq}\right)
\partial _{q}\partial _{r}q^{p}
=g^{pq}\partial _{p}\partial _{q}q^{a}-g^{aq}\partial _{q}\partial _{p}q^{p}
\end{equation}
which matches exactly with the result obtained in \cite{Padmanabhan2013a}. Applying the above result for the surface term in the \LL Lagrangian we obtain:
\begin{equation}
\pounds _{q}L_{\rm sur}=\pounds _{q}\partial _{a}\left(\sqrt{-g}V^{a}\right)
=\left(\pounds _{q}\partial _{a}\left[\sqrt{-g}V^{a}\right]\right)_{\rm std}
+\partial _{a}\left(\sqrt{-g}K^{a}\right)
\end{equation}
Since we have obtained Lie derivative of the surface term in the \LL Lagrangian we can compute Lie variation for the quadratic part. This leads to
\begin{align}\label{AppC:Eq.02}
\pounds _{q}\left(\sqrt{-g}L_{\rm quad}\right)&=\pounds _{q}\left(\sqrt{-g}L-L_{\rm sur}\right)
=\pounds _{q}\left(\sqrt{-g}L\right)-\left(\pounds _{q}L_{\rm sur}\right)_{\rm std}
-\partial _{a}\left(\sqrt{-g}K^{a}\right)
\nonumber
\\
&=\partial _{a}\left(\sqrt{-g}L_{\rm quad}q^{a}\right)-\partial _{a}\left(\sqrt{-g}K^{a}\right)
\end{align}
Also we have another useful identity which will appear frequently in this calculation given by
\begin{align}\label{Paper05_Final_Rev_01}
\pounds _{q}\mathcal{V}^{a}-q^{a}\partial _{c}\mathcal{V}^{c}
&=\left(\pounds _{q}\mathcal{V}^{a}\right)_{\rm std}-q^{a}\partial _{c}\mathcal{V}^{c}
+\sqrt{-g}K^{a}
\nonumber
\\
&=\partial _{c}\left(q^{c}\mathcal{V}^{a}-q^{a}\mathcal{V}^{c}\right)+\sqrt{-g}K^{a}
\end{align}
Note that even though $\pounds _{q}$ and $\delta$ commutes for $\Gamma ^{a}_{bc}$ (since $\delta$ commutes with partial derivatives) they do not commute for $\mathcal{V}^{a}$, since
\begin{align}
\delta \left(\pounds _{q}\mathcal{V}^{a}\right)=\delta \left(\pounds _{q}\mathcal{V}^{a}\right)_{\rm std}+\delta \left(\sqrt{-g}K^{a}\right)
\end{align}
Let us now try to obtain the Noether current corresponding to the quadratic Lagrangian. For that we start with the following decomposition:
\begin{align}
\pounds _{q}\left(\sqrt{-g}L_{\rm quad}\right)&=\pounds _{q}\left[\sqrt{-g}L-L_{\rm sur}\right]
\nonumber
\\
&=\sqrt{-g}E_{pq}\pounds _{q}g^{pq}
-m\partial _{c}\left(2\sqrt{-g}Q_{p}^{~qrc}\pounds _{q} \Gamma ^{p}_{qr}\right)
\nonumber
\\
&+\partial _{c}\left[2\sqrt{-g}Q_{p}^{~qrc}\pounds _{q} \Gamma ^{p}_{qr}
+\Gamma ^{p}_{qr}\pounds _{q} \left(2\sqrt{-g}Q_{p}^{~qrc}\right)\right]
\nonumber
\\
&=\partial _{a}\left[-2\sqrt{-g}E^{a}_{b}q^{b}+\Gamma ^{p}_{qr}\pounds _{q}\left(2\sqrt{-g}Q_{p}^{~qra}\right)
-2(m-1)\sqrt{-g}Q_{p}^{~qra}\pounds _{q} \Gamma ^{p}_{qr} \right]
\end{align}
Then from \eq{AppC:Eq.02} we arrive at the following conserved current:
\begin{align}
\sqrt{-g}J^{a}_{\rm quad}&=2\sqrt{-g}E^{a}_{b}q^{b}+\sqrt{-g}L_{\rm quad}q^{a}-\sqrt{-g}K^{a}
\nonumber
\\
&-\pounds _{q}\left(2\sqrt{-g}Q_{p}^{~qra}\right)\Gamma ^{p}_{qr}+2(m-1)\sqrt{-g}Q_{p}^{~qra}\pounds _{q} \Gamma ^{p}_{qr}
\nonumber
\\
&=\sqrt{-g}J^{a}-2m\sqrt{-g}Q_{p}^{~qra}\pounds _{q}\Gamma ^{p}_{qr}-q^{a}\sqrt{-g}L+\sqrt{-g}L_{\rm quad}q^{a}-\sqrt{-g}K^{a}
\nonumber
\\
&-\pounds _{q}\left(2\sqrt{-g}Q_{p}^{~qra}\right)\Gamma ^{p}_{qr}+2(m-1)\sqrt{-g}Q_{p}^{~qra}\pounds _{q} \Gamma ^{p}_{qr}
\nonumber
\\
&=\sqrt{-g}J^{a}-q^{a}L_{\rm sur}-\sqrt{-g}K^{a}-\pounds _{q}\left(2\sqrt{-g}Q_{p}^{~qra}\Gamma ^{p}_{qr}\right)
\nonumber
\\
&=\sqrt{-g}J^{a}-q^{a}\partial _{c}\left(\sqrt{-g}V^{c}\right)-\sqrt{-g}K^{a}+\pounds _{q}\left(\sqrt{-g}V^{a}\right)
\nonumber
\\
&=\sqrt{-g}J^{a}+\partial _{c}\left(\sqrt{-g}q^{c}V^{a}-\sqrt{-g}q^{a}V^{c}\right)
\end{align}
This ultimately leads to the following expression on using \eq{Paper05_Final_Rev_01}
\begin{equation}
J^{a}_{\rm quad}=J^{a}-2\nabla _{b}\left(\Gamma ^{p}_{qr}Q_{p}^{~qr[a}q^{b]}\right)
\end{equation}
This automatically shows that the Noether current and associated Noether potential corresponding to the surface Lagrangian is,
\begin{equation}
J^{a}_{\rm sur}=\nabla _{b}\left(q^{a}V^{b}-q^{b}V^{a}\right);\qquad
J^{ab}_{\rm sur}=q^{a}V^{b}-q^{b}V^{a}=2\left(q^{b}Q_{p}^{~qra}\Gamma ^{p}_{qr}
-q^{a}Q_{p}^{~qrb}\Gamma ^{p}_{qr}\right)
\end{equation}
Thus variation of the Noether current for the quadratic term leads to:
\begin{align}
\delta \left(\sqrt{-g}J^{a}_{\rm quad}\right)&=\delta \left(\sqrt{-g}J^{a}\right)
-\partial _{b}\left[\delta \Gamma ^{p}_{qr}\left(2\sqrt{-g}Q_{p}^{~qr[a}\right)q^{b]}
+\Gamma ^{p}_{qr}\delta \left(2\sqrt{-g}Q_{p}^{~qr[a}\right)q^{b]}\right]
\nonumber
\\
&=\sqrt{-g}E_{pq}\delta g^{pq}q^{a}+\sqrt{-g}\omega ^{a}+m\partial _{b}\left[\left(2\sqrt{-g}Q_{p}^{~qr[a}\right)q^{b]}
\delta \Gamma ^{p}_{qr}\right]
\nonumber
\\
&+2q^{b}\delta \left( \sqrt{-g}E^{a}_{b}\right)
-\partial _{b}\left[\delta \Gamma ^{p}_{qr}\left(2\sqrt{-g}Q_{p}^{~qr[a}\right)q^{b]}
+\Gamma ^{p}_{qr}\delta \left(2\sqrt{-g}Q_{p}^{~qr[a}\right)q^{b]}\right]
\nonumber
\\
&=\sqrt{-g}\omega ^{a}-\partial _{b}\left[\Gamma ^{p}_{qr}\delta\left(2\sqrt{-g}Q_{p}^{~qr[a}\right)q^{b]}\right]
+(m-1)\partial _{b}\left[\left(2\sqrt{-g}Q_{p}^{~qr[a}\right)q^{b]}
\delta \Gamma ^{p}_{qr}\right]+E^{a}
\end{align}
where the equation of motion terms $E^{a}$ can be given by
\begin{equation}
E^{a}=\sqrt{-g}E_{pq}\delta g^{pq}q^{a}+2q^{b}\delta \left( \sqrt{-g}E^{a}_{b}\right)
\end{equation}
Let us start with the definition for gravitational momentum in \LL gravity, which can be presented as follows:
\begin{align}
-\sqrt{-g}P^{a}&=2\sqrt{-g}P_{p}^{~qra}\pounds _{q}\Gamma ^{p}_{qr}+\sqrt{-g}Lq^{a}
\nonumber
\\
&=\pounds _{q}\left(2\sqrt{-g}P_{p}^{~qra}\Gamma ^{p}_{qr}\right)-m\Gamma ^{p}_{qr}\pounds _{q}\left(2\sqrt{-g}Q_{p}^{~qra}\right)+q^{a}\sqrt{-g}L_{\rm quad}-q^{a}\partial _{c}\left(2\sqrt{-g}Q_{p}^{~qrc}\Gamma ^{p}_{qr}\right)
\nonumber
\\
&=-P^{a}_{\rm quad}+\pounds _{q}\left(2\sqrt{-g}P_{p}^{~qra}\Gamma ^{p}_{qr}\right)-q^{a}\partial _{c}\left(2\sqrt{-g}Q_{p}^{~qrc}\Gamma ^{p}_{qr}\right)
\end{align}
where we have defined the quadratic momentum as:
\begin{equation}
P^{a}_{\rm quad}=\Gamma ^{p}_{qr}\pounds _{q}\left(2\sqrt{-g}P_{p}^{~qra}\right)
-q^{a}\sqrt{-g}L_{\rm quad}
\end{equation}
Then we can write the quadratic momentum using $\mathcal{V}^{a}$ as:
\begin{equation}
P^{a}_{\rm quad}=\sqrt{-g}P^{a}-m\pounds _{q}\mathcal{V}^{a}+q^{a}\partial _{c}\mathcal{V}^{c}
\end{equation}
Then using the result for Lie variation of $\mathcal{V}^{a}$ the quadratic momentum can be written as:
\begin{align}
P^{a}_{\rm quad}&=\sqrt{-g}P^{a}-m\left[\partial _{c}
\left(q^{c}\mathcal{V}^{a}-q^{a}\mathcal{V}^{c}\right)+\sqrt{-g}K^{a}
+q^{a}\partial _{c}\mathcal{V}^{c} \right]+q^{a}\partial _{c}\mathcal{V}^{c}
\nonumber
\\
&=\sqrt{-g}P^{a}-m\partial _{c}\left(q^{c}\mathcal{V}^{a}-q^{a}\mathcal{V}^{c}\right)
-m\sqrt{-g}K^{a}-(m-1)q^{a}\partial _{c}\mathcal{V}^{c}
\end{align}
Then under variations we obtain on-shell:
\begin{align}
\delta P^{a}_{\rm quad}&=\delta \left(\sqrt{-g}P^{a}\right)-m\delta \left(\sqrt{-g}K^{a}\right)+m\partial _{c}\Big[\delta \left(2\sqrt{-g}Q_{p}^{~qr[a}\right)q^{c]}\Gamma ^{p}_{qr}
\nonumber
\\
&+\left(2\sqrt{-g}Q_{p}^{~qr[a}\right)q^{c]}\delta \Gamma ^{p}_{qr}\Big]+(m-1)q^{a}\partial _{c}\left[\delta \left(2\sqrt{-g}Q_{p}^{~qrc}\right)\Gamma ^{p}_{qr}+\left(2\sqrt{-g}Q_{p}^{~qrc}\right)\delta \Gamma ^{p}_{qr}\right]
\nonumber
\\
&=-\sqrt{-g}\omega ^{a}-m\delta \left(\sqrt{-g}K^{a}\right)-m\partial _{c}\left[\left(2\sqrt{-g}Q_{p}^{~qr[a}\right)q^{c]}\delta \Gamma ^{p}_{qr}\right]
\nonumber
\\
&+m\partial _{c}\Big[\delta \left(2\sqrt{-g}Q_{p}^{~qr[a}\right)q^{c]}\Gamma ^{p}_{qr}+\left(2\sqrt{-g}Q_{p}^{~qr[a}\right)q^{c]}\delta \Gamma ^{p}_{qr}\Big]
\nonumber
\\
&+(m-1)q^{a}\partial _{c}\left[\delta \left(2\sqrt{-g}Q_{p}^{~qrc}\right)\Gamma ^{p}_{qr}+\left(2\sqrt{-g}Q_{p}^{~qrc}\right)\delta \Gamma ^{p}_{qr}\right]
\nonumber
\\
&=-\sqrt{-g}\omega ^{a}-m\delta \left(\sqrt{-g}K^{a}\right)
+m\partial _{c}\left[\delta \left(2\sqrt{-g}Q_{p}^{~qr[a}\right)q^{c]}\Gamma ^{p}_{qr}\right]
\nonumber
\\
&+(m-1)q^{a}\partial _{c}\left[\delta \left(2\sqrt{-g}Q_{p}^{~qrc}\right)\Gamma ^{p}_{qr}
+\left(2\sqrt{-g}Q_{p}^{~qrc}\right)\delta \Gamma ^{p}_{qr}\right]
\end{align}
These are the expressions used in the main text.

\subsubsection{Characterizing Null Surfaces}\label{Paper05:AppC:Null}

Let us now try to generalize the result for null surfaces to Lovelock gravity with the null vector $\ell _{a}$ such that $\ell ^{2}=0$ everywhere. For that purpose, we start with the combination:
\begin{align}
\mathcal{R}_{ab}\ell ^{a}\ell ^{b}&=R_{apqr}P_{b}^{~pqr}\ell ^{a}\ell ^{b}
\nonumber
\\
&=P^{bpqr}\ell _{b}\left(R_{apqr}\ell ^{a}\right)
\nonumber
\\
&=-P^{bpqr}\ell _{b}\left(\nabla _{q}\nabla _{r}\ell _{p}-\nabla _{r}\nabla _{q}\ell _{p}\right)
\nonumber
\\
&=-2P^{bpqr}\ell _{b}\nabla _{q}\nabla _{r}\ell _{p}
\nonumber
\\
&=\nabla _{q}\left(-2P^{bpqr}\ell _{b}\nabla _{r}\ell _{p}\right)-\mathcal{S}
\end{align}
where we have defined, the entropy density as:
\begin{align}
\mathcal{S}=-2P^{bpqr}\nabla _{q}\ell _{b}\nabla _{r}\ell _{p}
\end{align}
The Einstein-Hilbert limit can be obtained easily leading to:
\begin{align}
16\pi \mathcal{S}&=-\left(g^{bq}g^{pr}-g^{br}g^{pq}\right)\nabla _{q}\ell _{b}\nabla _{r}\ell _{p}
\nonumber
\\
&=\nabla _{a}\ell ^{b}\nabla _{b}\ell ^{a}-\left(\nabla _{i}\ell ^{i}\right)^{2}
\end{align}
as well as,
\begin{align}
32\pi P^{bpqr}\ell _{b}\nabla _{r}\ell _{p}&=\left(g^{bq}g^{pr}-g^{br}g^{pq}\right)\ell _{b}\nabla _{r}\ell _{p}
\nonumber
\\
&=\ell ^{q}\nabla _{r}\ell ^{r}-\ell ^{r}\nabla _{r}\ell ^{q}
\nonumber
\\
&=\Theta \ell ^{q}
\end{align}
However in \LL gravity, the combination, $2P^{bpqr}\ell _{b}\nabla _{r}\ell _{p}$ cannot be written as, $\phi \ell ^{q}$, for arbitrary $\phi$, since $2P^{bpqr}\ell _{q}\ell _{b}\nabla _{r}\ell _{p}\neq 0$. Next let us consider the object, $\ell _{a}J^{a}\left(\ell \right)$, for which we define, $\ell _{a}=A\nabla _{a}B$. Then in \LL gravity, we arrive at:
\begin{align}
\frac{1}{A}\ell _{a}J^{a}\left(\ell \right)&=\nabla _{b}\left[2P^{abcd}\ell _{a}\ell _{d}\nabla _{c}A\frac{1}{A^{2}} \right]
\nonumber
\\
&=\frac{1}{A}\nabla _{b}\left[2P^{abcd}\ell _{a}\ell _{d}\nabla _{c}\ln A \right]
-\frac{1}{A}\left[2P^{abcd}\ell _{a}\ell _{d}\nabla _{c}\ln A \nabla _{b}\ln A \right]
\end{align}
Let us now expand $\nabla _{c}\ln A$ in canonical null basis, such that we obtain,
\begin{align}
\nabla _{c}\ln A=-\kappa k_{c}+A\ell _{c}+B_{A}e^{A}_{c}
\end{align}
Note that we obtain, $\ell ^{c}\nabla _{c}\ln A=\kappa$. In a similar fashion, we can expand, $2P^{abcd}\ell _{a}\ell _{d}\nabla _{c}\ln A$ in the following manner,
\begin{align}
2P^{abcd}\ell _{a}\ell _{d}\nabla _{c}\ln A=P\ell ^{b}+Qk^{b}+R^{A}e^{b}_{A}
\end{align}
It is evident that $Q=-2P^{abcd}\ell _{a}\ell _{d}\ell _{b}\nabla _{c}\ln A=0$, due to antisymmetry of $P^{abcd}$ in the first two indices. Then it turns out that, 
\begin{align}
P=-2P^{abcd}\ell _{a}k_{b}\ell _{d}\nabla _{c}\ln A\equiv \mathcal{K}
\end{align}
It is obvious from the Einstein-Hilbert limit, that
\begin{align}
16\pi \mathcal{K}&=-\left(g^{ac}g^{bd}-g^{ad}g^{bc}\right)\ell _{a}k_{b}\ell _{d}\nabla _{c}\ln A
\nonumber
\\
&=-\left(g^{ac}g^{bd}-g^{ad}g^{bc}\right)\ell _{a}k_{b}\ell _{d}\left(-\kappa k_{c}+B_{A}e^{A}_{c}\right)
=\kappa
\end{align}
Also in the expansion for, $\nabla _{c}\ln A$, we obtain, $B_{A}$ is completely arbitrary. Then we can use $B_{A}$ such that the following relation: $B_{Q}P^{abcd}\ell _{a}\ell _{d}e^{A}_{b}e^{Q}_{c}=\kappa P^{abcd}\ell _{a}k_{c}\ell _{d}e^{A}_{b}$ is satisfied, such that, $R^{A}$ vanishes. Thus we obtain,
\begin{align}
2P^{abcd}\ell _{a}\ell _{d}\nabla _{c}\ln A=\mathcal{K}\ell ^{b}
\end{align}
Again, we get,
\begin{align}
2P^{abcd}\ell _{a}\ell _{d}\nabla _{c}\ln A \nabla _{b}\ln A=\mathcal{K}\ell ^{b}\nabla _{b}\ln A
=\kappa \mathcal{K}
\end{align}
Thus, finally we arrive at the following result:
\begin{align}
\ell _{a}J^{a}\left(\ell \right)=\nabla _{a}\left(\mathcal{K}\ell ^{a}\right)-\kappa \mathcal{K}
\end{align}
Again by considering derivative on the null surface, we obtain,
\begin{align}
D_{a}\left(\mathcal{K}\ell ^{a}\right)&=\left(g^{ab}+\ell ^{a}k^{b}+\ell ^{b}k^{a}\right)
\nabla _{a}\left(\mathcal{K}\ell _{b}\right)
\nonumber
\\
&=\nabla _{a}\left(\mathcal{K}\ell ^{a}\right)+k^{b}\ell ^{a}\nabla _{a}\left(\mathcal{K}\ell _{b}\right)
\nonumber
\\
&=\nabla _{a}\left(\mathcal{K}\ell ^{a}\right)-\kappa \mathcal{K}-\ell ^{a}\nabla _{a}\mathcal{K}
\end{align}
Thus the Noether current contraction can also be written as:
\begin{align}
\ell _{a}J^{a}\left(\ell \right)=D_{a}\left(\mathcal{K}\ell ^{a}\right)+\dfrac{d\mathcal{K}}{d\lambda}
\end{align} 

\providecommand{\href}[2]{#2}\begingroup\raggedright\endgroup

\begin{thebibliography}{10}

\bibitem{Lanczos1932}C. Lanczos, \textit{Z.Phys.} \textbf{73}, 147 (1932); C. Lanczos, \textit{Ann. of Math.} \textbf{39}, 842 (1938).

\bibitem{Lovelock1971}D. Lovelock, \textit{J. Math. Phys.} \textbf{12}, 498 (1971).

\bibitem{Padmanabhan2013b} T. Padmanabhan and D. Kothawala, \textit{Phys. Rept.} \textbf{531}, 115 (2013).

\bibitem{Padmanabhan2002} T. Padmanabhan, \textit{Class. Quant. Grav.} \textbf{19}, 5837 (2002) arXiv:0204019.

\bibitem{Padmanabhan2005a} T. Padmanabhan, \textit{Phys. Rept.} \textbf{406}, 49 (2005) arXiv:0311036.

\bibitem{Padmanabhan2010a} T. Padmanabhan, \textit{Rept. Prog. Phys.} \textbf{73}, 046901 (2010) arXiv:0911.5004.

\bibitem{Cai2005} R.G. Cai and S.P. Kim, \textit{J. High Energy Phys.} \textbf{0502}, 050 (2005) arXiv:hep-th/0501055.

\bibitem{Akbar2006} M. Akbar and R.G. Cai, \textit{Phys. Lett. B} \textbf{635}, 7 (2006) arXiv:hep-th/0602156.

\bibitem{Akbar2009} M. Akbar and M. Jamil, arXiv:0911.2556 (2009).

\bibitem{Akbar2007} M. Akbar \textit{Chin. Phys. Lett.} \textbf{24} 1158 (2007); M. Akbar and A.A. Siddiqui 
\textit{Phys. Lett. B} \textbf{656} 217 (2007).

\bibitem{Cai2008a} R.G. Cai, L.M. Cao and Y.P. Hu  JHEP 0808:090 (2008); M. Akbar and R.G. Cai \textit{Phys. Rev. D} \textbf{75} 084003 (2007).

\bibitem{Cai2007} R.G. Cai and L.M. Cao \textit{Nucl. Phys. B} \textbf{785} 135 (2007); A. Sheykhi, B. Wang and
R.G. Cai \textit{Nucl. Phys. B} \textbf{779} 1 (2007); A. Sheykhi, B. Wang and R.G. Cai \textit{Phys. Rev. D} \textbf{76} 023515 (2007); R.G. Cai \textit{Prog. Theor. Phys. Suppl.} \textbf{172} 100 (2008); X.H. Ge \textit{Phys. Lett. B} \textbf{651} 49 (2007).

\bibitem{Gong2007} Y. Gong and A. Wang \textit{Phys. Rev. Lett.} \textbf{99} 211301 (2007).

\bibitem{Wu2007} S.F. Wu, G.H. Yang and P.M. Zhang arXiv:0710.5394; S.F. Wu, B. Wang and G.H. Yang, 
\textit{Nucl. Phys. B} \textbf{799} 330 (2008); S.F. Wu et. al. \textit{Class. Quant. Grav.} \textbf{25} 235018
(2008); T. Zhu, J.R. Ren and S.F. Mo arXiv:0805.1162; M. Akbar \textit{Chin. Phys.
Lett.} \textbf{25} 4199-4202 (2008).

\bibitem{Padmanabhan2007b} D. Kothawala, S. Sarkar and T. Padmanabhan, \textit{Phys. Lett. B} \textbf{652}, 338 (2007).

\bibitem{Chakraborty2010} S. Chakraborty, R. Biswas and N. Mazumdar, \textit{Nuovo Cimento} \textbf{B125}, 1209 (2011) arXiv:1006.1169.

\bibitem{Chakraborty2009} N. Mazumdar and S. Chakraborty, \textit{Class. Quant. Grav.} \textbf{26}, 195016 (2009).

\bibitem{Padmanabhan2006a} A. Paranjape, S. Sarkar and T. Padmanabhan, \textit{Phys. Rev. D} \textbf{74}, 104015 (2006) arXiv:hep-th/0607240.

\bibitem{Padmanabhan2009} D. Kothawala and T. Padmanabhan, \textit{Phys. Rev. D} \textbf{79}, 104020 (2009) arXiv:0904.0215.

\bibitem{Bekenstein1973}  J.D. Bekenstein, \textit{Phys. Rev. D} \textbf{7}, 2333 (1973).

\bibitem{Bekenstein1974} J.D. Bekenstein, \textit{Phys. Rev. D} \textbf{9}, 3292 (1974).

\bibitem{Hawking1975} S. Hawking, \textit{Commun. Math. Phys.} \textbf{43}, 199 (1975).

\bibitem{Davies1976} P. C. W. Davies, S. A. Fulling, and W. G. Unruh, \textit{Phys. Rev. D} \textbf{13}, 2720 (1976).

\bibitem{Unruh1976} W. G. Unruh, \textit{Phys. Rev. D} \textbf{14}, 870 (1976).

\bibitem{Hawking1977} G. Gibbons and S.W. Hawking, \textit{Phys. Rev. D} \textbf{15}, 2752 (1977).

\bibitem{Jacobson1995} T. Jacobson, \textit{Phys. Rev. Lett.} \textbf{75}, 1260 (1995).

\bibitem{Padmanabhan2010b} T. Padmanabhan, \textit{Gravitation: Foundation and Frontiers}, Cambridge University Press, Cambridge, UK (2010).

\bibitem{Wald2001} R.M. Wald, \textit{Liv. Rev. Relt.} \textbf{4}, 6 (2001) arXiv:9912119.

\bibitem{Padmanabhan2008} T. Padmanabhan, \textit{Gen. Relt. Grav.} \textbf{40}, 529 (2008) arXiv:0705.2533.

\bibitem{Padmanabhan2002b} T. Padmanabhan, \textit{Mod. Phys. Lett. A} \textit{17}, 1147 (2002) arXiv:hep-th/0205278.

\bibitem{Padmanabhan2002c} T. Padmanabhan, \textit{Gen. Relt. Grav.} \textbf{34}, 2029 (2002) arXiv:gr-qc/0205090.

\bibitem{Padmanabhan2006c} A. Mukhopadhyay and T. Padmanabhan, \textit{Phys. Rev. D} \textbf{74}, 124023 (2006) arXiv:hep-th/0608120.

\bibitem{Kolekar2010} S. Kolekar and T. Padmanabhan, \textit{Phys. Rev. D} \textbf{82}, 024036 (2010) 
arXiv:1005.0619.

\bibitem{Kolekar2012b} S. Kolekar, D. Kothawala and T. Padmanabhan, \textit{Phys. Rev. D} \textbf{85}, 064031 (2012) arXiv:1111.0973.

\bibitem{Padmanabhan2011a} T. Padmanabhan, \textit{Phys. Rev. D} \textbf{83}, 044048 (2011) arXiv:1012.0119. 

\bibitem{Kolekar2012a} S. Kolekar and T. Padmanabhan, \textit{Phys. Rev. D} \textbf{85}, 024004 (2012) arXiv:1109.5353.

\bibitem{Damour1982} T. Damour, \textit{Surface Effects in Black Hole Physics}, Proceedings of the second Marcel Grossmann Meeting on General Relativity (1982).

\bibitem{Padmanabhan2007} T. Padmanabhan and A. Paranjape, \textit{Phys. Rev. D} \textbf{75}, 064004 (2007) arXiv:gr-qc/0701003.

\bibitem{Krishna2013} K. Parattu, B.R. Majhi and T. Padmanabhan, \textit{Phys. Rev. D} \textbf{87}, 124011 (2013) arXiv:1303.1535.

\bibitem{Sumanta2014b}S.~Chakraborty and T.~Padmanabhan, ``{Geometrical variables with direct
  thermodynamic significance in Lanczos-Lovelock gravity},''
  \href{http://dx.doi.org/10.1103/PhysRevD.90.084021}{{\em Phys.Rev.}
  {\bfseries D90} no.~8, (2014) 084021},
\href{http://arxiv.org/abs/1408.4791}{{\ttfamily arXiv:1408.4791 [gr-qc]}}.

\bibitem{PadmanabhanH2013} H. Padmanabhan and T. Padmanabhan, \textit{Int. J. Mod. Phys. D} \textbf{22}, 1342001 (2013) arXiv:1302.3226.

\bibitem{PadmanabhanH2014} T. Padmanabhan and H. Padmanabhan, \textit{Int. J. Mod. Phys. D} \textbf{23}, 1430011 (2014).

\bibitem{Padmanabhan2013a} T. Padmanabhan, \textit{Gen. Rel. Grav} \textbf{46}, 1673 (2014) arXiv:1312.3253.

\bibitem{Sumanta2014a} S.~Chakraborty and T.~Padmanabhan, ``{Evolution of Spacetime arises due to the
  departure from Holographic Equipartition in all Lanczos-Lovelock Theories of
  Gravity},'' \href{http://dx.doi.org/10.1103/PhysRevD.90.124017}{{\em
  Phys.Rev.} {\bfseries D90} no.~12, (2014) 124017},
\href{http://arxiv.org/abs/1408.4679}{{\ttfamily arXiv:1408.4679 [gr-qc]}}.

\bibitem{Chakraborty2014c} B.~R. Majhi and S.~Chakraborty, ``{Anomalous effective action, Noether current,
  Virasoro algebra and Horizon entropy},''
  \href{http://dx.doi.org/10.1140/epjc/s10052-014-2867-6}{{\em Eur.Phys.J.}
  {\bfseries C74} (2014) 2867},
\href{http://arxiv.org/abs/1311.1324}{{\ttfamily arXiv:1311.1324 [gr-qc]}}.


\bibitem{Padmanabhan2012b} B.R. Majhi and T. Padmanabhan, \textit{Phys. Rev. D} \textbf{85}, 084040 (2012) 
arXiv:1111.1809.

\bibitem{Wald1993} R.M. Wald, \textit{Phys. Rev. D} \textbf{48}, R3427 (1993);\\
V. Iyer and R.M. Wald, \textit{Phys. Rev. D} \textbf{50}, 846 (1994);\\
R.M. Wald and A. Zoupas, \textit{Phys. Rev. D} \textbf{61}, 084027 (2000).

\bibitem{Strominger1996} A. Strominger and C. Vafa, \textit{Phys. Lett. B} \textbf{379}, 99 (1996).

\bibitem{Ashtekar1998} A. Ashtekar, J. Baez, A. Corichi and K. Krasnov, 
\textit{Phys. Rev. Lett.} \textbf{80}, 904 (1998);\\
J.M. Garcia-Islas, \textit{Class. Quant. Grav.} \textbf{25}, 245001 (2008).

\bibitem{Bombelli1986} L. Bombelli, R.K. Koul, J. Lee and R.D. Sorkin, 
\textit{Phys. Rev. D} \textbf{34}, 373 (1986).

\bibitem{Moncrief1983} V. Moncrief and J. Isenberg, \textit{Commun. Math. Phys.} \textbf{89}, 387 (1983).

\bibitem{Morales2008} E.M. Morales, \textit{On a Second Law of Black Hole Mechanics in a Higher Derivative Theory of Gravity} Ph.D Thesis, Gottingen University (2008).

\bibitem{Parattu2015a} K.~Parattu, S.~Chakraborty, B.~R. Majhi, and T.~Padmanabhan, ``{Null Surfaces:
  Counter-term for the Action Principle and the Characterization of the
  Gravitational Degrees of Freedom},''
\href{http://arxiv.org/abs/1501.01053}{{\ttfamily arXiv:1501.01053 [gr-qc]}}.

\bibitem{Dadhich2015} N. Dadhich and J.M. Pons, \textit{JHEP} \textit{1505}, 067 (2015) arXiv:1503.00974.

\bibitem{Dadhich2012} N. Dadhich and J.M. Pons, \textit{J. Math. Phys.} \textbf{54}, 102501 (2013) arXiv:1210.1109.

\bibitem{Dadhich2010} N. Dadhich, \textit{Pramana}, \textbf{74}, 875 (2010) arXiv:0802.3034.

\bibitem{Chakraborty2015a} S.~Chakraborty, K.~Parattu, and T.~Padmanabhan, ``{Gravitational Field
  equations near an Arbitrary Null Surface expressed as a Thermodynamic
  Identity},''
\href{http://arxiv.org/abs/1505.05297}{{\ttfamily arXiv:1505.05297 [gr-qc]}}.

\bibitem{Kothawala2010} D. Kothawala, \textit{Phys. Rev. D} {\bf 83}, 024026 (2011) arXiv:1010.2207.

\bibitem{Hayward1997} S.A. Hayward, \textit{Class. Quant. Grav.}  {\bf 15}, 3147 (1998) arXiv:gr-qc/9710089.

\bibitem{Kastor2009} D. Kastor, S. Ray and J. Traschen, \textit{Class. Quant. Grav.} {\bf 26}, 195011 (2009) arXiv:0904.2765.

\bibitem{Dolan2011} B.P. Dolan, \textit{Class. Quant. Grav.} {\bf 28}, 235017 (2011) arXiv:1106.6260.

\bibitem{Mann2012} D. Kubiznak and R.B. Mann, \textit{JHEP} {\bf 07(2012)}, 033 (2012) arXiv:1205.0559.

\bibitem{Mann2014} A.M. Frassino, D. Kubiznak, R.B. Mann and F. Simovic, \textit{JHEP} \textbf{09(2014)}, 080 (2014), arxiv:1406.7015.

\bibitem{Jacobson1993} T. Jacobson, R.C. Myres, \textit{Phys. Rev. Lett.} \textbf{70}, 3684 (1993).

\bibitem{Clunan2004} T. Clunan, S. Ross and D. Smith, \textit{Class. Quantum Grav.} \textbf{21}, 3447 (2004).

\bibitem{Julia1998} B. Julia and S. Silva, \textit{Class. Quant. Grav.} \textbf{15}, 2173 (1998) arXiv:gr-qc/9804029.

\bibitem{Regge1974} T. Regge and C. Teitelboim, \textit{Ann. Phys.} \textbf{88}, 286 (1974).

\bibitem{Carlip1999} S. Carlip, \textit{Class. Quant. Grav.} \textbf{16}, 3327 (1999) arXiv:gr-qc/9906126.

\end{thebibliography}
\end{document}